\documentclass[aps,reprint,onecolumn,amsmath,amssymb,longbibliography,notitlepage,raggedbottom]{revtex4-1}

\usepackage{graphicx}
\usepackage[caption=false]{subfig}
\usepackage{bm}
\usepackage{stmaryrd}
\usepackage{geometry} 
\usepackage{color}
\usepackage[normalem]{ulem}
\usepackage{xcolor}
\usepackage{ar}

\begin{document}

\preprint{APS/123-QED}

\title{On the use of eddy viscosity in resolvent analysis of turbulent channel flow}

\author{Sean Symon}
 \email{sean.symon@soton.ac.uk}
\affiliation{Aerodynamics and Flight Mechanics, Faculty of Engineering and Physical Sciences, University of Southampton, SO17 1BJ, United Kingdom}


\author{Anagha Madhusudanan}

\affiliation{Department of Applied Mathematics and Theoretical Physics, Centre for Mathematical Sciences,
University of Cambridge, Cambridge CB3 0WA, United Kingdom}

\author{Simon J. Illingworth}
\affiliation{Department of Mechanical Engineering, University of Melbourne, Parkville, VIC 3010, Australia}

\author{Ivan Marusic}
\affiliation{Department of Mechanical Engineering, University of Melbourne, Parkville, VIC 3010, Australia}

%

\date{\today}

\begin{abstract}

\vspace{0.5in}

The predictions of resolvent analysis for turbulent channel flow are evaluated for a friction Reynolds number of $Re_\tau = 550$. In addition to the standard resolvent operator where only the kinematic viscosity appears, a resolvent operator augmented with the Cess eddy viscosity profile is considered. Adding eddy viscosity significantly alters the low-rank behavior of the resolvent. Regardless of the wave speed selected, the eddy resolvent is low-rank for spanwise wavelengths of $\lambda^+_y = 80$ and $\lambda_y/h = 3.5$ in comparison to the standard resolvent whose low-rank behavior depends on the wave speed. The leading eddy resolvent modes are shown to have higher projections onto the leading mode from spectral proper orthogonal decomposition in comparison to standard resolvent modes. Neither analysis, however, reliably predicts the most energetic wave speed. The standard resolvent tends to overestimate it while the eddy resolvent underestimates it. For scales where the most energetic wave speed is underestimated, the eddy resolvent modes are energetic too close to the wall. The eddy resolvent does, however, correctly identify the most energetic wave speed and mode shapes for structures that are associated with the near-wall cycle or that are most energetic at $z/h = \pm 0.5$. It is argued that these types of structures are likely to be correctly predicted for any friction Reynolds number due to the inner and outer scaling of the Cess eddy viscosity profile. Finally, it is shown that the accuracy of eddy predictions relies on striking the right balance between positive and negative energy transfers. Even though the eddy viscosity primarily adds dissipation, its wall-normal gradient injects energy in the near-wall region, resulting in mode shapes that are ``attached'' to the wall. For some scales, however, the predicted positive energy transfer is too strong thus biasing structures towards the wall. The ability of the Cess eddy viscosity profile to model both positive and negative energy transfers suggests that it could be optimized for individual scales to provide better low-order models of turbulent channel flow. 

\end{abstract}

\maketitle

\section{Introduction}

The Navier-Stokes equations linearized around the mean (time-averaged) flow have been used to identify coherent structures in a variety of flows. In the resolvent-based approach of Ref.~\cite{McKeon10}, for example, the linearized Navier-Stokes equations are analyzed from an input-output perspective. The input is made up of nonlinear perturbation terms that are treated as an intrinsic forcing to the linear resolvent operator and the output is the perturbation velocity field. In the context of wall-bounded flows, resolvent analysis has been exploited for a variety of applications from finding reduced-order models of exact coherent states \cite{Sharma16} to predicting statistics of high-Reynolds-number turbulence \cite{Moarref13,Skouloudis21}. The ability of resolvent analysis to identify prominent linear mechanisms, moreover, has made it an attractive alternative to direct numerical simulation (DNS) for designing flow control strategies \cite{Luhar14,Nakashima17,Toedtli19}. 

From DNS and experimental studies of wall-bounded flows, it is known that the production of turbulent kinetic energy is mainly driven by the exchange of energy from the mean flow to the fluctuations. Resolvent analysis models well this exchange from the mean to the fluctuations, where the mean is assumed to be known \textit{a priori}.  Therefore, the term in the energy budget that the resolvent analysis captures most successfully is production \cite{Symon21,Jin21}. The resolvent operator also tends to be low-rank for energy-producing scales \cite{Moarref13}. The term that resolvent analysis captures the least accurately for an arbitrary scale is nonlinear transfer between scales. One way to model this term is to add an eddy viscosity to the resolvent operator \cite{Reynolds72, Pujals09, Hwang10, Morra19, Kuhn21, Morra21, Pickering21}. The eddy viscosity provides additional dissipation that removes energy from all scales. As such, it attempts to model this nonlinear transfer of energy from the large scales to the small scales. 

In terms of structures, many studies have observed good agreement between the structures predicted by resolvent analysis, and those found in DNS.  The leading resolvent mode, which is computed as the leading left singular vector of the resolvent operator, is the dominant structure predicted by resolvent analysis.  The dominant mode from DNS is computed as the leading mode from spectral proper orthogonal decomposition (SPOD) \cite{Lumley67,Picard00}.  In fact, if the nonlinear forcing is white in space and time, then resolvent and SPOD modes are theoretically equivalent \cite{Towne18}.  The nonlinear forcing, however, is not white in space and time \cite{Morra21,Nogeuria21,Amaral21}. As such, the role of eddy viscosity is to model the effect of the nonlinear forcing such that white in space and time forcing is sufficient to predict the correct structures. Reference~\cite{Morra19} showed that adding eddy viscosity improved predictions of coherent motions in turbulent channel flow at a friction Reynolds number of $Re_\tau = 1007$.  However, they only considered the streamwise velocity component of two of the energetic scales in the flow. Reference~\cite{Symon20} compared resolvent predictions with and without eddy viscosity for turbulent channel flow at $Re_\tau = 2003$. The study was limited, however, to temporal snapshots of the largest scales. The most comprehensive comparison between resolvent and SPOD modes has been presented by Ref.~\cite{Abreu20}, who computed the projection of the leading resolvent mode onto the leading SPOD mode for turbulent pipe flow at low Reynolds numbers. The authors noted good agreement, i.e. high projections between the resolvent and SPOD modes, for scales where the lift-up mechanism \cite{Ellingsen75,Landahl80} was active. They did not, however, consider the improvements that could be achieved by adding an eddy viscosity model as done by Ref.~\cite{Tissot21}.

A rigorous comparison of resolvent and SPOD modes across all scales and wave speeds, therefore, is needed for resolvent analysis with and without eddy viscosity. As such, one of the principal objectives of this article is to quantify the projections of the leading resolvent mode computed with and without eddy viscosity onto the leading SPOD mode. Previous work \cite{Symon20} suggests that analyzing the accuracy with which the resolvent model predicts the wave speed at which a structure convects is crucial when comparing the model predictions with DNS. This motivates the second objective of this work which is to perform a detailed analysis of the most amplified wave speeds identified by resolvent analysis with and without eddy viscosity and how they compare to the most energetic wave speed computed from DNS. The most energetic wave speed of a structure is also indicative of the wall-normal location where most of its energy is concentrated and, therefore, of the shape of the mode in the wall-normal direction. The final objective of this article is to provide insight into the types of nonlinear interactions that the eddy viscosity is attempting to model. As mentioned earlier, the eddy viscosity primarily adds extra dissipation to the energy balance, but its wall-normal gradient has been shown to add energy overall to a scale \cite{Symon21}. It is worth examining the wall-normal profiles of the transfers introduced by eddy viscosity to determine if they model the positive energy transfer in the near-wall region that has been observed in DNS for large scales \cite{Lee19,Kawata21,Doohan21, Hernandez21b}. 

The rest of the paper is organised as follows. Section~\ref{sec:methodology} presents the governing equations for channel flow and provides a brief overview of resolvent analysis, the Cess eddy viscosity profile, and SPOD. The details of the DNS for $Re_{\tau}  = 550$ are provided in Sec.~\ref{sec:DNS}. The predictions of resolvent and eddy analysis are evaluated in Sec.~\ref{sec:linear predictions} using scalar quantities. These include the ratio of energy contained in the first pair of singular values compared to the total energy, the projections of the leading resolvent modes onto the leading SPOD modes, and comparing the most amplified wave speed from resolvent analysis to the most energetic wave speed in DNS. Section~\ref{sec:mode shapes} provides a more detailed comparison of the mode shapes for select scales. The extent to which a constant eddy viscosity profile, i.e. one that does not depend on space, can successfully predict structures is also investigated in Sec.~\ref{sec:mode shapes}. The energy transfer processes introduced by eddy viscosity are discussed in Sec.~\ref{sec:energy}. The role of the eddy viscosity gradient, in particular, is examined and artificially manipulated to understand its influence on the mode shapes. Finally, conclusions and implications for optimizing a scale-dependent eddy viscosity are suggested in Sec.~\ref{sec:conclusions}.

\section{Methodology} \label{sec:methodology}

Section~\ref{sec:equations} describes the governing equations for plane Poiseuille flow and their non-dimensionalization. A brief overview of resolvent analysis is provided in Sec.~\ref{sec:resolvent analysis}. In Sec.~\ref{sec:Cess}, a modified resolvent operator, which includes the Cess eddy viscosity profile, is formulated. In order to assess the predictive capability of resolvent analysis with and without eddy viscosity, the leading modes are compared to SPOD modes, which are computed using the procedure summarized in Sec.~\ref{sec:SPOD}.

\subsection{Plane Poiseuille  flow equations} \label{sec:equations}

The non-dimensional Navier-Stokes equations for statistically steady, turbulent plane Poiseuille flow are
\begin{subequations} \label{eq:NSE}
	\begin{equation} \label{eq:Momentum}
	\frac{\partial \boldsymbol{u}}{\partial t} + \boldsymbol{u} \cdot \boldsymbol{\nabla} \boldsymbol{u}  = -\boldsymbol{\nabla} p + \frac{1}{Re_{\tau}}\boldsymbol{\nabla}^2\boldsymbol{u},	
	\end{equation} 
	\begin{equation}
	\boldsymbol{\nabla} \cdot \boldsymbol{u} = 0,
	\end{equation}
\end{subequations}
where $\boldsymbol{u}(\boldsymbol{x},t) = [u,v,w]^T$ is the velocity in the $x$ (streamwise), $y$ (spanwise), and $z$ (wall-normal) directions, $p(\boldsymbol{x},t)$ is the pressure, and $\boldsymbol{\nabla} = [\partial/\partial x,\partial/\partial y, \partial/\partial z]^T$. The friction Reynolds number $Re_{\tau} = u_\tau h /\nu$ is defined in terms of the friction velocity $u_{\tau}$, channel half height $h$, and kinematic viscosity $\nu$. Periodic boundary conditions are applied in the streamwise and spanwise directions and no-slip boundary conditions are imposed at the walls. The velocities are non-dimensionalized by $u_\tau$, the spatial variables by $h$, and the pressure by $\rho u_{\tau}^2$ where $\rho$ is the density of the fluid. A `$+$' superscript denotes spatial variables that have been normalized by the viscous length scale $\nu/u_{\tau}$. 

\subsection{Resolvent analysis} \label{sec:resolvent analysis}

Equation~(\ref{eq:NSE}) is Reynolds-decomposed  leading to the following equations for the fluctuations:
\begin{subequations} \label{eq:input-output}
\begin{equation}
\frac{\partial \boldsymbol{u}'}{\partial t} + \boldsymbol{U} \cdot \boldsymbol{\nabla} \boldsymbol{u}' + \boldsymbol{u}' \cdot \boldsymbol{\nabla} \boldsymbol{U} + \boldsymbol{\nabla} p' - \frac{1}{Re_{\tau}}\boldsymbol{\nabla}^2 \boldsymbol{u}' = -\boldsymbol{u}'\cdot \boldsymbol{\nabla} \boldsymbol{u}' + \overline{\boldsymbol{u}' \cdot \boldsymbol{\nabla} \boldsymbol{u}'} = \boldsymbol{f}',
\end{equation}
\begin{equation}
\boldsymbol{\nabla} \cdot \boldsymbol{u}' = 0,
\end{equation}
\end{subequations}
where $(\overline{\cdot})$ and $(\cdot)'$ denote a time-average and fluctuation, respectively. The mean velocity profile $\boldsymbol{U} = [U(z),0,0]^T$ is assumed to be known \textit{a priori} from DNS. Equation~(\ref{eq:input-output}) is written such that all linear terms appear on the left-hand side. The nonlinear terms on the right-hand side are lumped together as a forcing $\boldsymbol{f}'$. Equation~(\ref{eq:input-output}) is Laplace-transformed in time and Fourier-transformed in the homogeneous directions $x$ and $y$
\begin{equation} \label{eq:Laplace transform}
\hat{\boldsymbol{u}}(k_x,k_y,s) = \frac{1}{(2 \pi)^3} \int_{-\infty}^{\infty}\int_{-\infty}^{\infty} \int_{-\infty}^{\infty} \boldsymbol{u}'(x,y,z,t) e^{st-ik_xx-ik_yy}dxdydt.
\end{equation}
Upon integration of Eq.~(\ref{eq:Laplace transform}), we set $s=i\omega$ to consider the frequency response $\hat{\boldsymbol{u}}(\boldsymbol{k})$ where $(\hat{\cdot})$ denotes the Fourier-transformed coefficient and the wavenumber triplet $\boldsymbol{k} = (k_x,k_y,\omega)$ consists of streamwise wavenumber $k_x$, spanwise wavenumber $k_y$, and temporal frequency $\omega$. The equivalent wavelengths in the streamwise and spanwise directions are $\lambda_x = 2\pi/k_x$ and $\lambda_y = 2\pi/k_y$. The wavenumbers are non-dimensionalized by $(1/h)$ and the wavelengths by $h$. 

Equation~(\ref{eq:Laplace transform}) is substituted into Eq.~(\ref{eq:input-output}) and rearranged into state-space form \citep{Jovanovic05}
\begin{subequations} \label{eq:OSSQ}
	\begin{equation} 
	i\omega \hat{\boldsymbol{q}}(\boldsymbol{k}) = \boldsymbol{A}(k_x,k_y)\hat{\boldsymbol{q}}(\boldsymbol{k})  + \boldsymbol{B}(k_x,k_y)\hat{\boldsymbol{f}}(\boldsymbol{k}) ,
	\end{equation} 
	\begin{equation}
	\hat{\boldsymbol{u}}(\boldsymbol{k})  = \boldsymbol{C}(k_x,k_y)\hat{\boldsymbol{q}}(\boldsymbol{k}) , 
	\end{equation}
\end{subequations}
where the state $\hat{\boldsymbol{q}}$ consists of the wall-normal velocity $\hat{w}$ and wall-normal vorticity $\hat{\eta} = ik_y\hat{u} - ik_x\hat{v}$. The matrices $\boldsymbol{A}$, $\boldsymbol{B}$, and $\boldsymbol{C}$ are the discretized forms of the linearized Navier-Stokes operator, the forcing operator and the output operator, respectively, and are defined in the Appendix. It is worth noting that $\boldsymbol{A}$, $\boldsymbol{B}$, and $\boldsymbol{C}$ are independent of $\omega$ but are functions of the wavenumber pair $(k_x,k_y)$ under consideration. For the sake of brevity, this dependence is omitted for the rest of the paper.

Equation~(\ref{eq:OSSQ}) is recast into input-output form
\begin{equation}
\hat{\boldsymbol{u}}(\boldsymbol{k}) = \boldsymbol{C}(i\omega \boldsymbol{I}- \boldsymbol{A})^{-1} \hat{\boldsymbol{f}}(\boldsymbol{k}) = \mathcal{H}(\boldsymbol{k}) \hat{\boldsymbol{f}}(\boldsymbol{k}),
\end{equation}
where $\mathcal{H}(\boldsymbol{k})$ is a linear operator called the resolvent that relates the input forcing $\hat{\boldsymbol{f}}(\boldsymbol{k})$ to the output velocity $\hat{\boldsymbol{u}}(\boldsymbol{k})$. Even if $\hat{\boldsymbol{f}}(\boldsymbol{k})$ is unknown, the resolvent operator can be characterized by the singular value decomposition
\begin{equation} \label{eq:SVD}
\mathcal{H}(\boldsymbol{k}) = \hat{\boldsymbol{\Psi}} (\boldsymbol{k}) \boldsymbol{\Sigma}(\boldsymbol{k}) \hat{\boldsymbol{\Phi}}^*(\boldsymbol{k}),
\end{equation} 
where $\hat{\boldsymbol{\Psi}}(\boldsymbol{k}) = [\hat{\boldsymbol{\psi}}_1(\boldsymbol{k}), \hat{\boldsymbol{\psi}}_2(\boldsymbol{k}), \cdots, \hat{\boldsymbol{\psi}}_p(\boldsymbol{k})]$ are the resolvent modes, which form an orthogonal basis for velocity, and  $\hat{\boldsymbol{\Phi}}(\boldsymbol{k}) = [\hat{\boldsymbol{\phi}}_1(\boldsymbol{k}),\hat{\boldsymbol{\phi}}_2(\boldsymbol{k}), \cdots, \hat{\boldsymbol{\phi}}_p(\boldsymbol{k})]$ are the resolvent forcing modes which form an orthogonal basis for the nonlinear forcing. $\boldsymbol{\Sigma}(\boldsymbol{k})$ is a diagonal matrix that ranks the $p$th structure by its gain $\sigma_p(\boldsymbol{k})$ using an inner product that is proportional to its kinetic energy, i.e. $\left< \hat{\boldsymbol{\psi}},\hat{\boldsymbol{\psi}} \right> = \int_{-h}^h \hat{\boldsymbol{\psi}}^* \cdot \hat{\boldsymbol{\psi}}dz$. The structure $\hat{\boldsymbol{\psi}}_1(\boldsymbol{k})$ is, therefore, referred to as the optimal or leading resolvent mode and is the most amplified response of the linear dynamics contained in the resolvent. 

The true velocity field from experiments or DNS can be expressed as a weighted sum of resolvent modes
\begin{equation} \label{eq:weighted sum}
\hat{\boldsymbol{u}}(\boldsymbol{k}) = \sum_{p = 1}^N \hat{\boldsymbol{\psi}}_p(\boldsymbol{k}) \sigma_p(\boldsymbol{k})  \chi_p(\boldsymbol{k}),
\end{equation}
where $\chi_p(\boldsymbol{k})$ is the projection of $\hat{\boldsymbol{\phi}}_p(\boldsymbol{k})$ onto $\hat{\boldsymbol{f}}(\boldsymbol{k})$, i.e. 
\begin{equation} \label{eq:nonlinear projections}
\chi_p(\boldsymbol{k}) = \left<\hat{\boldsymbol{f}}(\boldsymbol{k}),\hat{\boldsymbol{\phi}}_p(\boldsymbol{k}) \right>. 
\end{equation}
It can be noted from Eqs.~(\ref{eq:weighted sum}) and~(\ref{eq:nonlinear projections}) that if $\hat{\boldsymbol{f}}(\boldsymbol{k})$ is white noise or projects equally onto the resolvent forcing modes, then the contribution of a resolvent mode $\hat{\boldsymbol{\psi}}_p$ in reconstructing the velocity field according to Eq.~(\ref{eq:weighted sum}) is solely dependent on the associated singular value $\sigma_p$. Moreover, if $\sigma_1 \gg \sigma_2$, then it is often argued that the velocity response can be well-approximated by the leading resolvent response mode $\hat{\boldsymbol{\psi}}_1$ alone. 

\subsection{Cess eddy viscosity model} \label{sec:Cess}

It has been shown, however, that the nonlinear forcing may have little to no overlap with the leading forcing mode \cite{Morra21, Symon21, Barthel21} resulting in $\chi_1 \ll \chi_{p \neq 1}$. This nonalignment stems from the fact that the resolvent does not model well the inter-scale nonlinear transfer. To address this shortcoming, an eddy viscosity can be added to the linearized Navier-Stokes equations after performing a triple decomposition of the total velocity field $\tilde{\boldsymbol{u}}$ into a mean component $\boldsymbol{U}$, coherent motions $\boldsymbol{u}$, and incoherent fluctuations $\boldsymbol{u}'$ \cite{Reynolds72}. A new set of equations govern the coherent velocity and pressure:
\begin{equation} \label{eq:triple}
\frac{\partial \boldsymbol{u}}{\partial t} + \boldsymbol{U} \cdot \boldsymbol{\nabla} \boldsymbol{u} + \boldsymbol{u} \cdot \boldsymbol{\nabla} \boldsymbol{U} + \boldsymbol{\nabla} p + \nabla \cdot \left[\nu_T(\boldsymbol{\nabla} \boldsymbol{u} + \boldsymbol{\nabla} \boldsymbol{u}^T) \right] = \boldsymbol{d},
\end{equation}
where $\nu_T(z)$ is the total effective viscosity and $\boldsymbol{d} =  - \boldsymbol{u} \cdot \boldsymbol{\nabla u}  + \overline{\boldsymbol{u} \cdot \boldsymbol{\nabla} \boldsymbol{u}} $ is the disturbance term. Similar to Refs.~\cite{Reynolds67} and \cite{Hwang10}, the Cess \cite{Cess58} eddy viscosity profile 
\begin{equation} \label{eq:Cess}
\nu_T(z) = \frac{\nu}{2} \left(1 + \left[ \frac{\kappa}{3} (1-z^2) (1+2z^2)( 1-\text{exp} (|z-1|Re_{\tau}/A) \right]^2 \right)^{1/2} + \frac{\nu}{2},
\end{equation}
is employed in this study. The constants $\kappa = 0.426$ and $A = 25.4$ are chosen based on a least-squares fit to experimental mean velocity profiles at $Re_{
\tau} = 2000$ \cite{delAlamo06}. Even though the Reynolds number in this study is lower than the Reynolds number for which the fit was performed, it has been verified that the results are not sensitive to the values of these constants.  

As done in Sec.~\ref{sec:resolvent analysis}, Eq.~(\ref{eq:triple}) is Fourier-transformed in time and the homogeneous directions to obtain an input-output relationship between the velocity and disturbance fields
\begin{equation}
\hat{\boldsymbol{u}}(\boldsymbol{k}) = \mathcal{H}^e(\boldsymbol{k})\hat{\boldsymbol{d}}(\boldsymbol{k}),
\end{equation}
where $\mathcal{H}^e(\boldsymbol{k})$ is a modified resolvent operator. Its singular value decomposition can be written as
\begin{equation} \label{eq:eddy svd}
\mathcal{H}^e(\boldsymbol{k}) = \hat{\boldsymbol{\Psi}}^e(\boldsymbol{k})\boldsymbol{\Sigma}^e(\boldsymbol{k})\hat{\boldsymbol{\Phi}}^{*,e}(\boldsymbol{k}).
\end{equation}
The superscript $e$ differentiates the eddy resolvent modes, henceforth referred to as eddy modes, and singular values from their standard resolvent counterparts in Eq.~(\ref{eq:SVD}). The interpretation of each term in the decomposition is similar to Eq.~(\ref{eq:SVD}) in that $\hat{\Psi}^e(\boldsymbol{k})$ consists of orthogonal basis functions for the velocity field and the diagonal matrix $\boldsymbol{\Sigma}^e(\boldsymbol{k})$ ranks the $p$th structure by its gain using an inner product proportional to its kinetic energy. The matrix  $\hat{\Phi}^e(\boldsymbol{k})$, on the other hand, contains orthogonal basis functions for the disturbance field $\hat{\boldsymbol{d} }(\boldsymbol{k})$ which is less interpretable than $\hat{\boldsymbol{f}}(\boldsymbol{k})$. Despite this drawback, the addition of eddy viscosity is expected to partially model the effect of $\hat{\boldsymbol{f}}(\boldsymbol{k})$ and thus improve the efficiency of eddy modes as a basis for the velocity field. 

\subsection{Spectral proper orthogonal decomposition} \label{sec:SPOD}

The efficiency of resolvent and eddy modes as a basis for the velocity field can be assessed by projecting them onto SPOD modes, which are computed directly from data. The SPOD modes are computed with the same procedure described in Ref.~\cite{Towne18} so only a brief summary is presented here. Using Welch's method \citep{Welch67}, the DNS data for a particular $(k_x,k_y)$ are divided into overlapping segments containing 512 snapshots with 75\% overlap. Each segment is Fourier-transformed in time and the Fourier modes for a specific frequency $\omega$ can be arranged into the new data matrix
\begin{equation}
\hat{\boldsymbol{Q}}(\boldsymbol{k}) = \left[\begin{array}{cccc} \hat{\boldsymbol{q}}_{\omega}^{(1)} & \hat{\boldsymbol{q}}_{\omega}^{(2)} & \cdots & \hat{\boldsymbol{q}}_{\omega}^{(s)}  \end{array} \right] \in \mathbb{C}^{m \times s}, 
\end{equation}
where $m$ represents the number of states and $s$ the number of segements. The cross-spectral density matrix for a specific wavenumber triplet $\hat{\boldsymbol{S}}(\boldsymbol{k})$ is 
\begin{equation} \label{eq:CSD}
\hat{\boldsymbol{S}}(\boldsymbol{k}) = \hat{\boldsymbol{Q}}(\boldsymbol{k})\hat{\boldsymbol{Q}}^*(\boldsymbol{k})  .
\end{equation} 
The SPOD eigenvectors (or modes) $\hat{\boldsymbol{V}}(\boldsymbol{k})$ and eigenvalues $\boldsymbol{\Lambda}(\boldsymbol{k})$ can be obtained by performing an eigenvalue decomposition of the cross-spectral density matrix
\begin{equation}
\hat{\boldsymbol{S}}(\boldsymbol{k})\hat{\boldsymbol{V}}(\boldsymbol{k}) = \hat{\boldsymbol{V}}(\boldsymbol{k})\boldsymbol{\Lambda}(\boldsymbol{k}).
\end{equation}

\section{DNS dataset} \label{sec:DNS}

\begin{table}
	\begin{center}
		\def~{\hphantom{0}}
		\begin{tabular}{ccccccccccc}
			$Re_{\tau}$  &   $L_x$ & $L_y$ & $L_z$ & $N_x$ & $N_y$ & $N_z$ &  $\Delta x^+$ & $\Delta y^+$ & $\Delta z^+_{min}$ & $\Delta z^+_{max}$   \\[3pt]
			 550  & $2\pi$ & $\pi$ & $2h$ & 256 & 256 & 201 & 13.5 & 6.75 & $6.79 \times 10^{-2}$ & 8.64 \\
		\end{tabular}
		\caption{Channel flow DNS parameters.}
		\label{tab:parameters}
	\end{center}
\end{table}

\begin{figure} 
\centering
\includegraphics[scale=0.4]{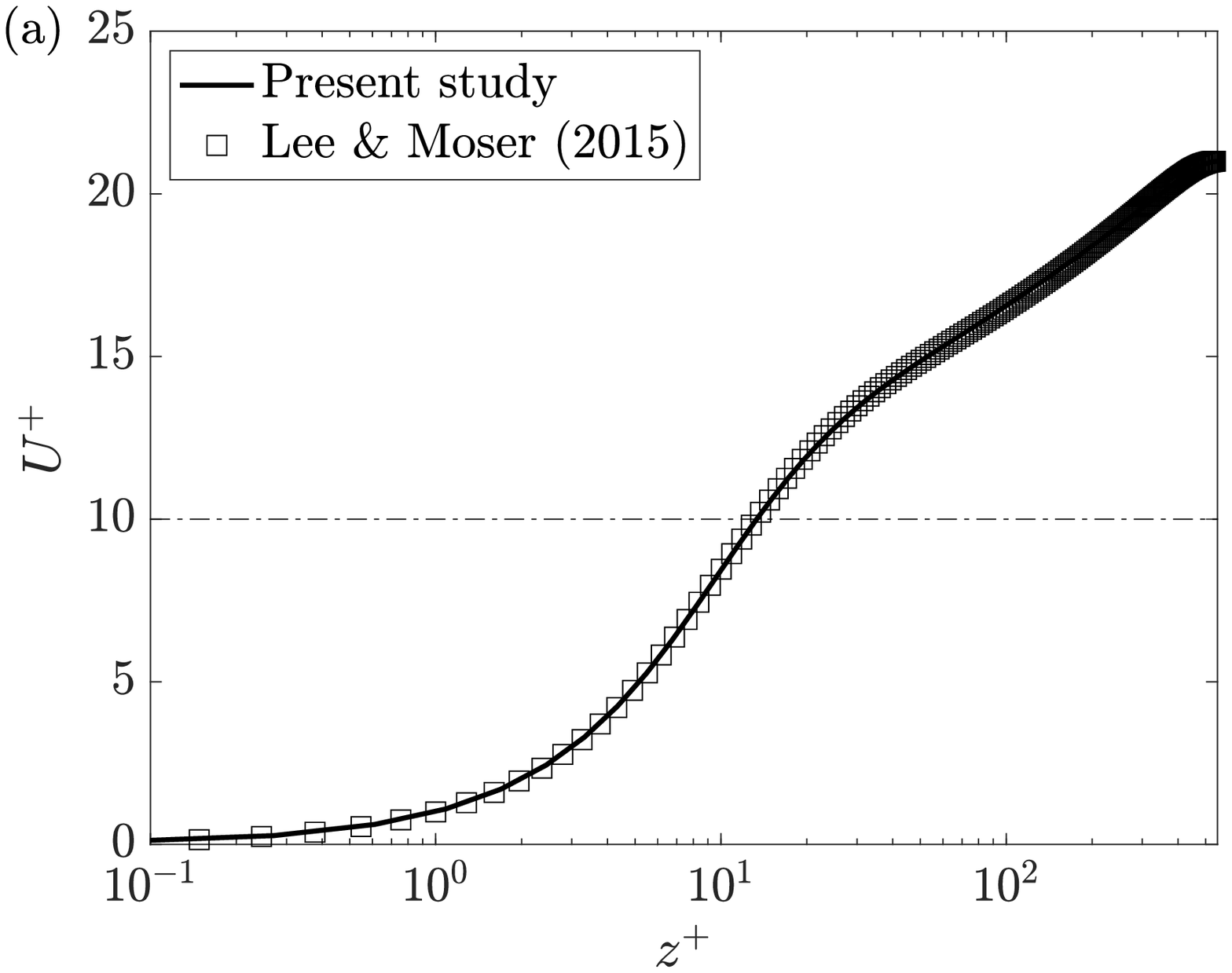}
\includegraphics[scale=0.4]{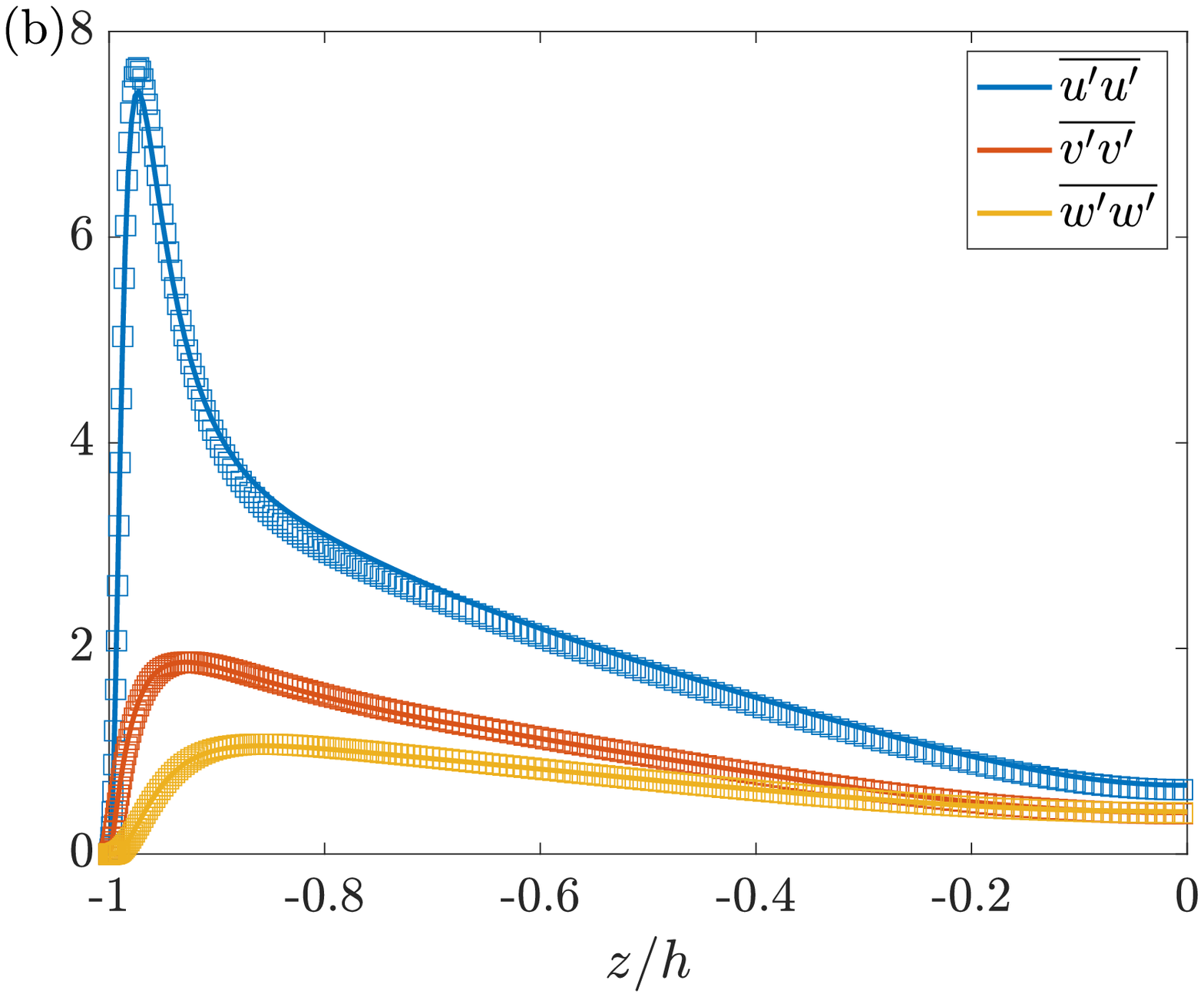}
\caption{(a) Mean velocity profile and (b) Reynolds stress profiles from the present study (solid lines) and Ref.~\cite{Lee15} (open squares).} \label{fig:DNS profiles}
\end{figure}

A DNS of channel flow at $Re_{\tau} = 550$ is performed using the ChannelFlow pseudo-spectral code \cite{Gibson19}. Table~\ref{tab:parameters} summarizes the parameters of the simulation which was solved on a domain with dimensions $2\pi \times \pi \times 2h$ in the streamwise ($L_x$), spanwise ($L_y$) and wall-normal ($L_z$) directions. There are $N_x = N_y = 256$ equally spaced points in the streamwise and spanwise directions and $N_y = 201$ points in the wall-normal direction on a Chebyshev grid. Periodic boundary conditions are employed in the streamwise and spanwise directions while no-slip boundary conditions are enforced on the channel walls. Further details on the mesh discretization $(\Delta x^+,\Delta y^+, \Delta z^+_{min}, \Delta z^+_{\max})$ are presented in Table~\ref{tab:parameters}. The mean velocity and Reynolds stress profiles are presented in Figs.~\ref{fig:DNS profiles}(a) and (b), respectively. All profiles show good agreement with the DNS results from Ref.~\cite{Lee15} despite the smaller computational box in this study. 

SPOD is performed on a database of 6784 snapshots at a time resolution of $\Delta t = 0.2$. The data are divided into equal segments containing 512 snapshots with an overlap of 75\% resulting in 50 blocks. The cross-spectral density matrices in Eq.~(\ref{eq:CSD}) are computed using Welch's method with a Hamming window. The SPOD modes and their respective energies for a desired frequency are obtained from the eigenvectors and eigenvalues, respectively, of the cross-spectral density matrices. 

\section{Resolvent and eddy analysis predictions} \label{sec:linear predictions}

In this section, the predictions from resolvent and eddy analysis, i.e. the resolvent supplemented with an eddy viscosity model, are compared to DNS data using several measures. To begin with, the low-rank behavior of the resolvent is compared to the turbulent kinetic energy spectrum of DNS in Sec.~\ref{sec:lrmaps}. Next, the projection of the leading resolvent and eddy modes onto the dominant SPOD mode is computed in Sec.~\ref{sec:projections} for the most energetic wavenumber pairs. Structures convecting at a range of wave speeds $c^+ = \omega/k_x$ are considered, and projections are computed independently for each wave speed. In Sec.~\ref{sec:frequency analysis}, the most energetic wave speed for a range of $(k_x,k_y)$ wavenumber pairs in DNS is compared to the most amplified wave speed identified by resolvent and eddy analysis. This motivates the selection of specific wavenumber triplets to examine in greater detail in Sec.~\ref{sec:mode shapes} in which the SPOD, resolvent, and eddy mode shapes are compared directly.  

\subsection{Low-rank maps} \label{sec:lrmaps}

An important aspect of the resolvent operator is its rank or, more specifically, the ratio of the total energy that is captured by the leading resolvent modes alone. Reference~\cite{Moarref13} was able to show that for a turbulent channel flow at $Re_{\tau} = 2003$, in the region of the $(k_x,k_y)$ space where the real flow is energetic (as seen from experiments or DNS), the standard resolvent operator tends to be low-rank. With this observation in mind, in this section the low-rank maps from resolvent and eddy analysis are compared to the turbulent kinetic energy spectra from DNS (data provided by Refs.~~\cite{delAlamo03} and \cite{delAlamo04}). As noted by Ref.~\cite{Moarref13}, the symmetry of the channel leads to the resolvent singular values coming in (approximately) equal pairs. One mode in each pair is symmetric with respect to the channel centerline while the other mode is anti-symmetric. The low-rank behavior of the resolvent for the channel can therefore be studied by computing the ratio of the dominant pair of singular values to the sum of all singular values, i.e.
\begin{equation}
\mathcal{R}(\boldsymbol{k}) = \frac{\sigma_1^2(\boldsymbol{k})+\sigma_2^2(\boldsymbol{k})}{\sum_p \sigma_p^2(\boldsymbol{k})}.
\end{equation}
Since $\mathcal{R}$ depends on a wavenumber triplet, $\mathcal{R}$ is computed across $(\lambda_x,\lambda_y)$ for a fixed wave speed in order to facilitate visualization. 

\begin{figure} 
\centering
\includegraphics[trim=0cm 0cm 0cm 0cm, clip=true, scale=0.4]{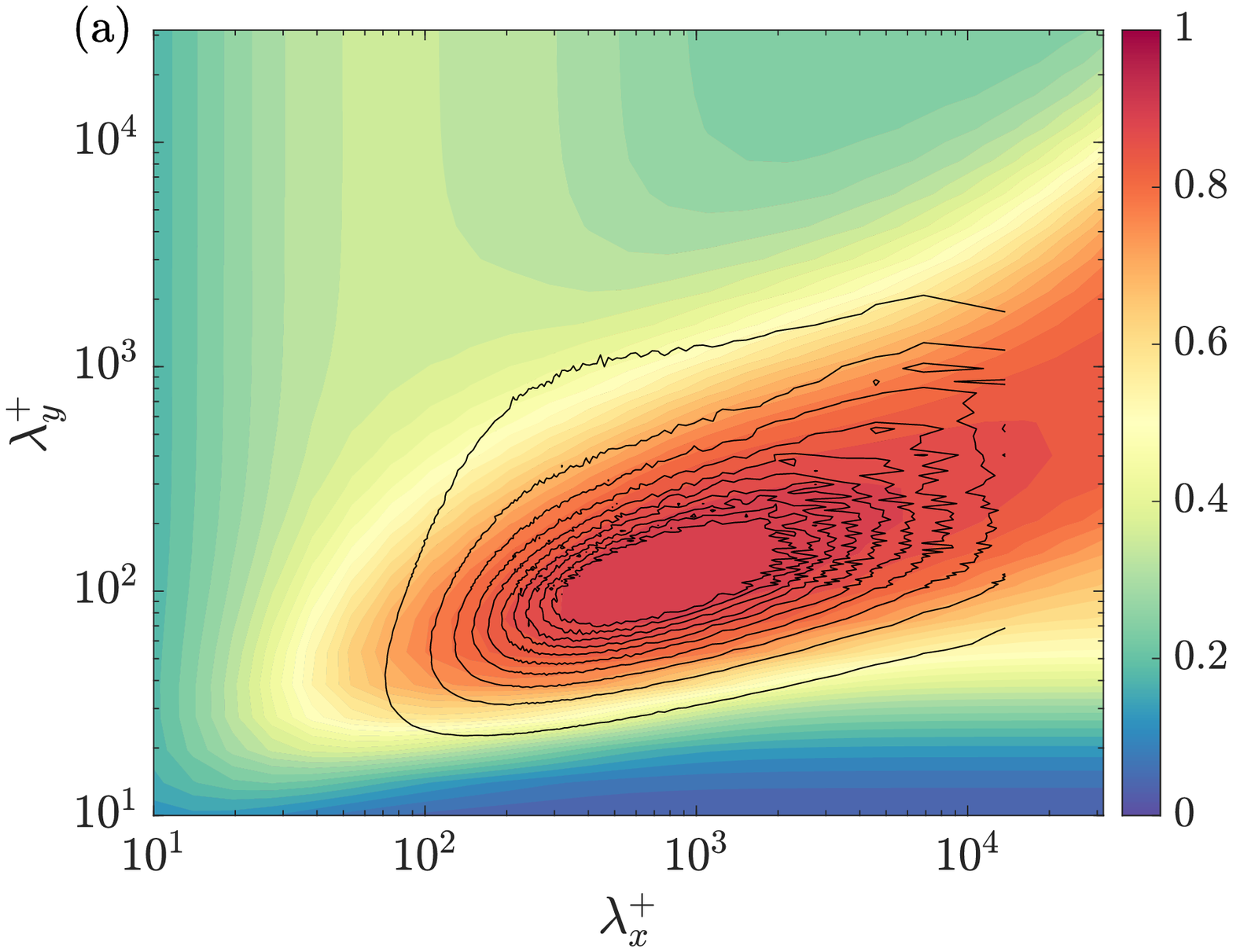}
\includegraphics[trim=0cm 0cm 0cm 0cm, clip=true, scale=0.4]{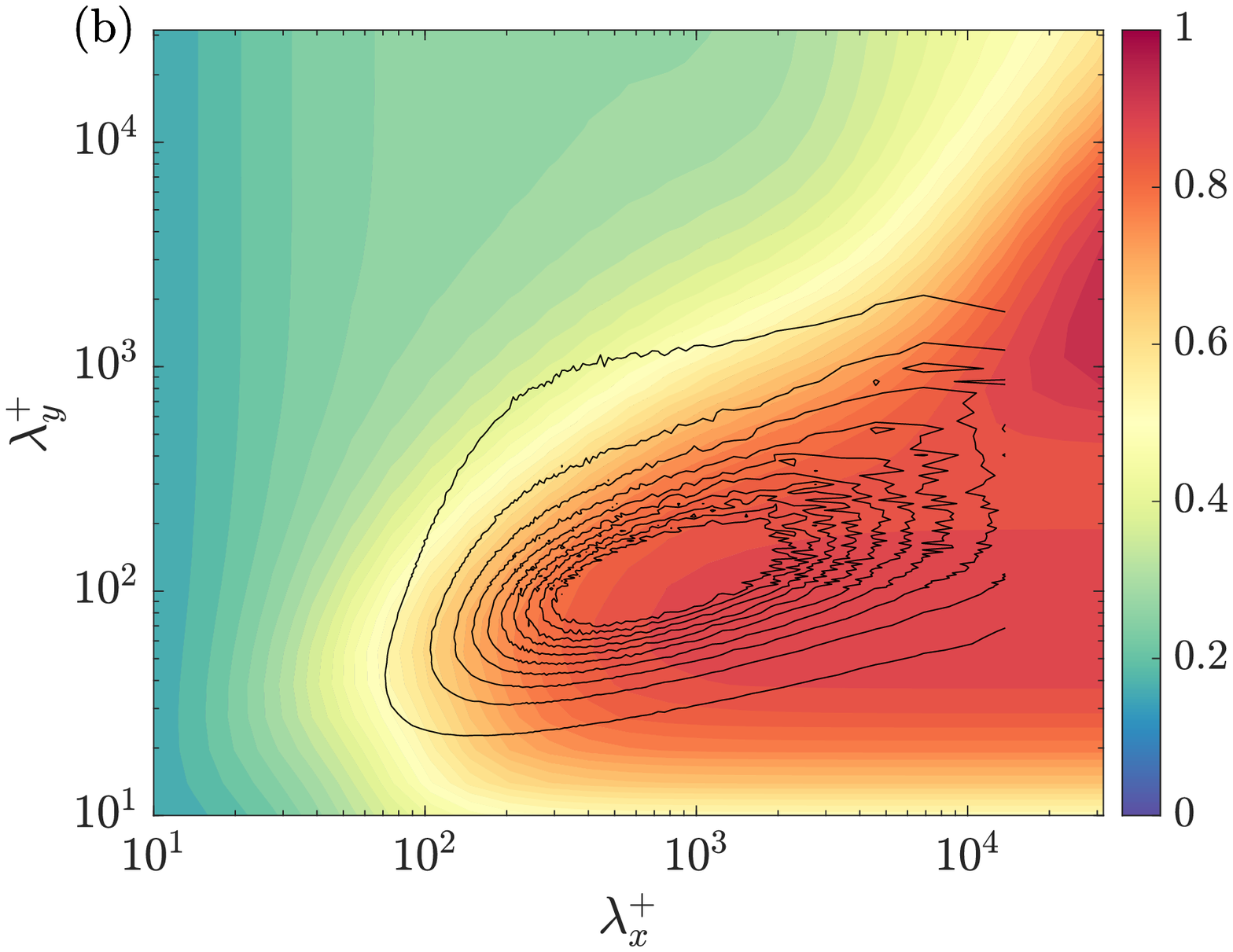}
\caption{Low-rank maps for (a) resolvent and (b) eddy analysis for a fixed wave speed of $c^+ = 10$. Contours of the turbulent kinetic energy spectrum at $z^+ = 15$ from Refs.~\cite{delAlamo03} and \cite{delAlamo04} are denoted in black.} \label{fig:low rank map1}
\end{figure}

The low-rank maps for $c^+ = 10$ are presented in Figs.~\ref{fig:low rank map1}(a) and (b) for resolvent and eddy analysis, respectively. The wall-normal location corresponding to $U^+ = 10$ is approximately $z^+ = 15$ as indicated by the dash-dotted line in Fig.~\ref{fig:DNS profiles}(a). The colours in the figure represent the value of $\mathcal{R}(\boldsymbol{k})$, while the black contour lines represent the streamwise turbulent kinetic energy spectrum from DNS at $z^+\approx15$ \cite{delAlamo03, delAlamo04}. Figure~\ref{fig:low rank map1}(a) shows that there is very good agreement between the scales with the most energy in DNS and the scales at which the resolvent is low-rank. This is to be expected since the energetic structures in the flow tend to arise from linear amplification mechanisms that are identified well by resolvent analysis \cite{Jovanovic05, McKeon10}. The eddy low-rank map in Fig.~\ref{fig:low rank map1}(b) has two main differences from its resolvent counterpart in Fig.~\ref{fig:low rank map1}(a). The first is that there is less agreement between the low-rank map and the energy spectra. The most energetic scales from DNS with spanwise wavelengths of $\lambda_y^+ \approx 100$, for example, do not coincide with the peak in the eddy low-rank map which occurs at $\lambda_y^+ \approx 80$. The second is that there are two peaks in the low-rank map for the eddy analysis whereas there is only one for resolvent analysis. The two peaks from the eddy operator occur at $\lambda_y^+ = 80$ and $\lambda_y = 3.5h$, wavelengths that have been previously observed by Refs.~\cite{delAlamo06} and \cite{Hwang10} in the context of transient growth and harmonic forcing analyses, respectively. It can be observed that there is some energy from larger scales present at this wall-normal location as attested to by the closed contour of kinetic energy centered on $(\lambda_x,\lambda_y)= (14h,2h)$ or $(\lambda_x^+,\lambda_y^+) = (7700,1100)$. Nevertheless, the Reynolds number of this study is too low to conclude that the second peak around $\lambda_y = 3.5h$ is indicative of the footprint of large-scales at the wall.

\begin{figure} 
\centering
\includegraphics[trim=0cm 0cm 0cm 0cm, clip=true, scale=0.4]{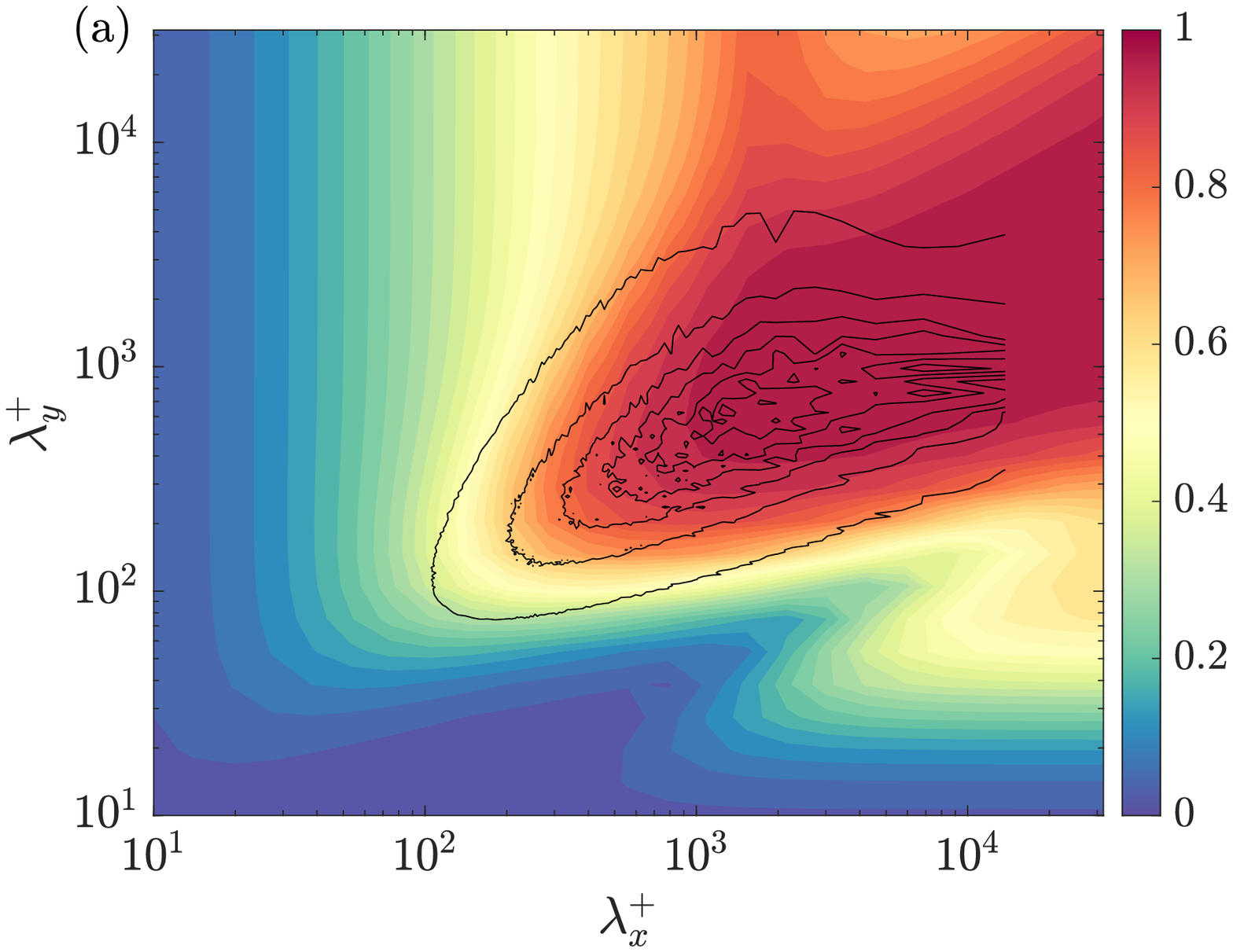}
\includegraphics[trim=0cm 0cm 0cm 0cm, clip=true, scale=0.4]{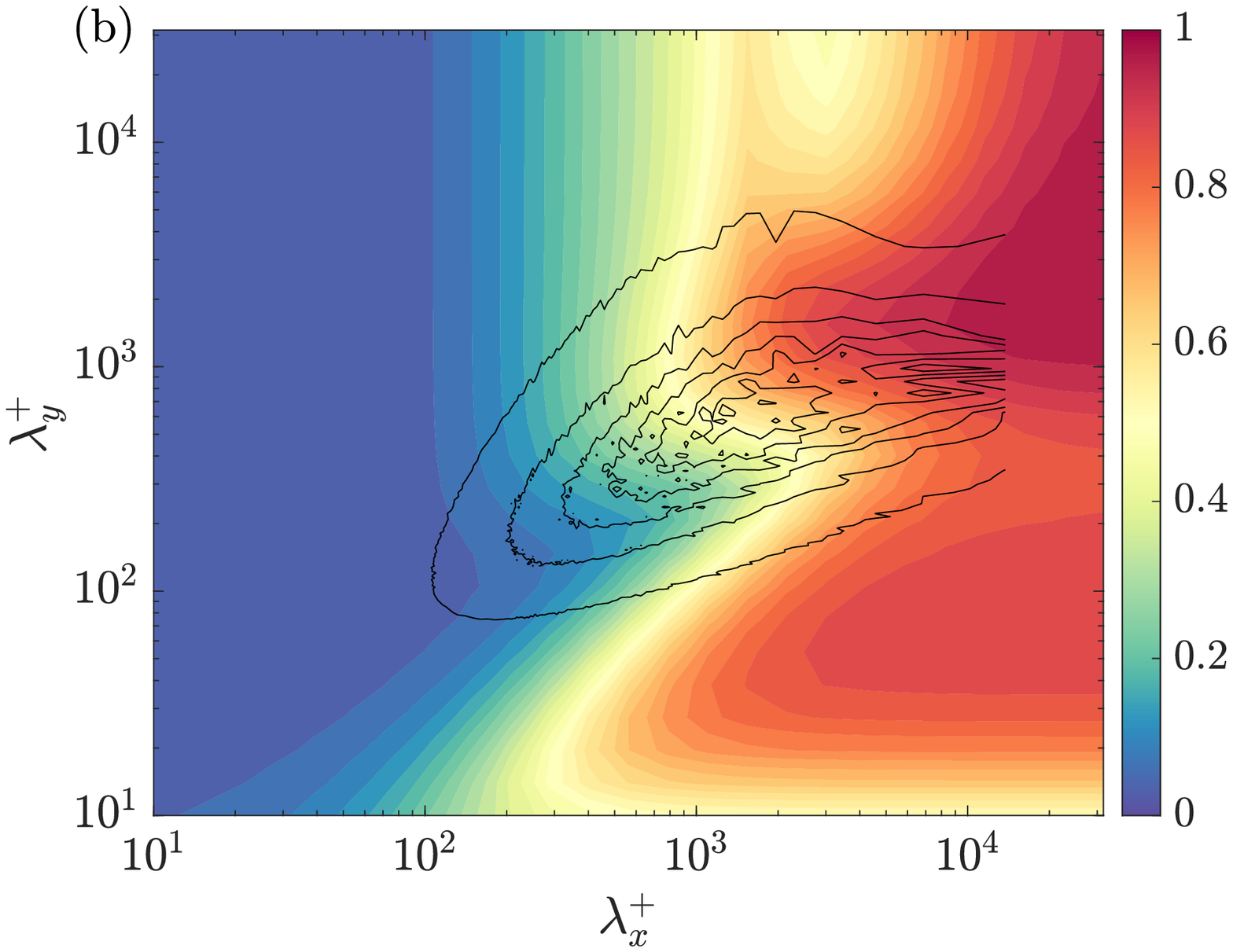}
\caption{Low-rank maps for (a) resolvent and (b) eddy analysis for a fixed wave speed of $c^+ = 18.9$. Contours of the turbulent kinetic energy spectrum at $z^+ = -h/2$ from Refs.~\cite{delAlamo03} and \cite{delAlamo04} are denoted in black.} \label{fig:low rank map2}
\end{figure}

The low-rank maps for $c^+ = 18.9$ are shown in Fig.~\ref{fig:low rank map2}. Similar to the previous wave speed, there is good agreement between the energy spectrum from DNS and the low-rank map for resolvent analysis in Fig.~\ref{fig:low rank map2}(a). The agreement between the DNS and low-rank map for eddy analysis in Fig.~\ref{fig:low rank map2}(b), on the other hand, is worse than it was for the previous wave speed. The only area of slight agreement occurs for scales with the longest streamwise wavelengths from DNS. Perhaps the most remarkable aspect of the low-rank map in Fig.~\ref{fig:low rank map2}(b) is that it closely resembles the low-rank map in Fig.~\ref{fig:low rank map1}(b). The resolvent low-rank maps, on the other hand, are influenced significantly by the wave speed. It can be concluded that the wave speed does not have a significant impact on the eddy low-rank maps. For both wave speeds considered, there are two spanwise wavelengths for which the eddy operator is low-rank. The only difference is that $\mathcal{R}$ for $c^+ = 18.9$ decreases for the smaller spanwise wavelength and increases for the larger spanwise wavelength relative to $\mathcal{R}$ for $c^+ = 10$. 

It can be remarked that the addition of eddy viscosity significantly distorts the linear mechanisms identified by resolvent analysis. As such, eddy analysis is less successful in identifying the energetic scales for a specified wave speed. These results, notwithstanding, do not quantify the accuracy of resolvent or eddy analysis in predicting flow structures in turbulent channel flow. It will be seen that the two peak spanwise wavelengths identified by the low-rank maps provide a valuable clue in identifying the types of structures that eddy analysis predicts with good accuracy.  

\subsection{Projection of resolvent modes onto SPOD modes} \label{sec:projections}

The objective of this section is to quantify the accuracy of resolvent and eddy analysis by projecting the leading SPOD mode $\hat{\boldsymbol{v}}_1(\boldsymbol{k})$ from DNS onto the leading resolvent $\hat{\boldsymbol{\psi}}_1(\boldsymbol{k})$ and eddy $\hat{\boldsymbol{\psi}}^e_1(\boldsymbol{k})$ modes. Similar analyses have been performed by Refs.~\cite{Abreu20} and \cite{Pickering21} for turbulent pipe flow and turbulent jets, respectively. To account for the pairing of resolvent and eddy modes, the leading SPOD mode is projected onto both the first and second resolvent/eddy modes
\begin{subequations} \label{eq:projection gamma}
\begin{equation}
\gamma(\boldsymbol{k}) = \sqrt{\left( \frac{ \left< \hat{\boldsymbol{v}}_1(\boldsymbol{k}),\hat{\boldsymbol{\psi}}_1(\boldsymbol{k}) \right> }{\| \hat{\boldsymbol{v}}_1(\boldsymbol{k})  \| \cdot \| \hat{\boldsymbol{\psi}}_1(\boldsymbol{k}) \|}  \right) ^2  + \left( \frac{ \left< \hat{\boldsymbol{v}}_1(\boldsymbol{k}),\hat{\boldsymbol{\psi}}_2(\boldsymbol{k}) \right> }{\| \hat{\boldsymbol{v}}_1(\boldsymbol{k})  \| \cdot \| \hat{\boldsymbol{\psi}}_2(\boldsymbol{k}) \|}  \right) ^2  },
\end{equation}
\begin{equation}
\gamma^e(\boldsymbol{k}) = \sqrt{\left( \frac{ \left< \hat{\boldsymbol{v}}_1(\boldsymbol{k}),\hat{\boldsymbol{\psi}}^e_1(\boldsymbol{k}) \right> }{\| \hat{\boldsymbol{v}}_1(\boldsymbol{k})  \| \cdot \| \hat{\boldsymbol{\psi}}^e_1(\boldsymbol{k}) \|}  \right) ^2  + \left( \frac{ \left< \hat{\boldsymbol{v}}_1(\boldsymbol{k}),\hat{\boldsymbol{\psi}}^e_2(\boldsymbol{k}) \right> }{\| \hat{\boldsymbol{v}}_1(\boldsymbol{k})  \| \cdot \| \hat{\boldsymbol{\psi}}^e_2(\boldsymbol{k}) \|}  \right) ^2  }.
\end{equation}
\end{subequations}
Both projection coefficients $\gamma(\boldsymbol{k})$ and $\gamma^e(\boldsymbol{k})$ have a maximum value of unity, which indicates perfect alignment between SPOD and resolvent/eddy modes. A value of zero indicates that the mode shapes are orthogonal.

\begin{figure} 
\centering
\includegraphics[trim=4cm 0.5cm 0cm 1.5cm, clip=true, scale=0.4]{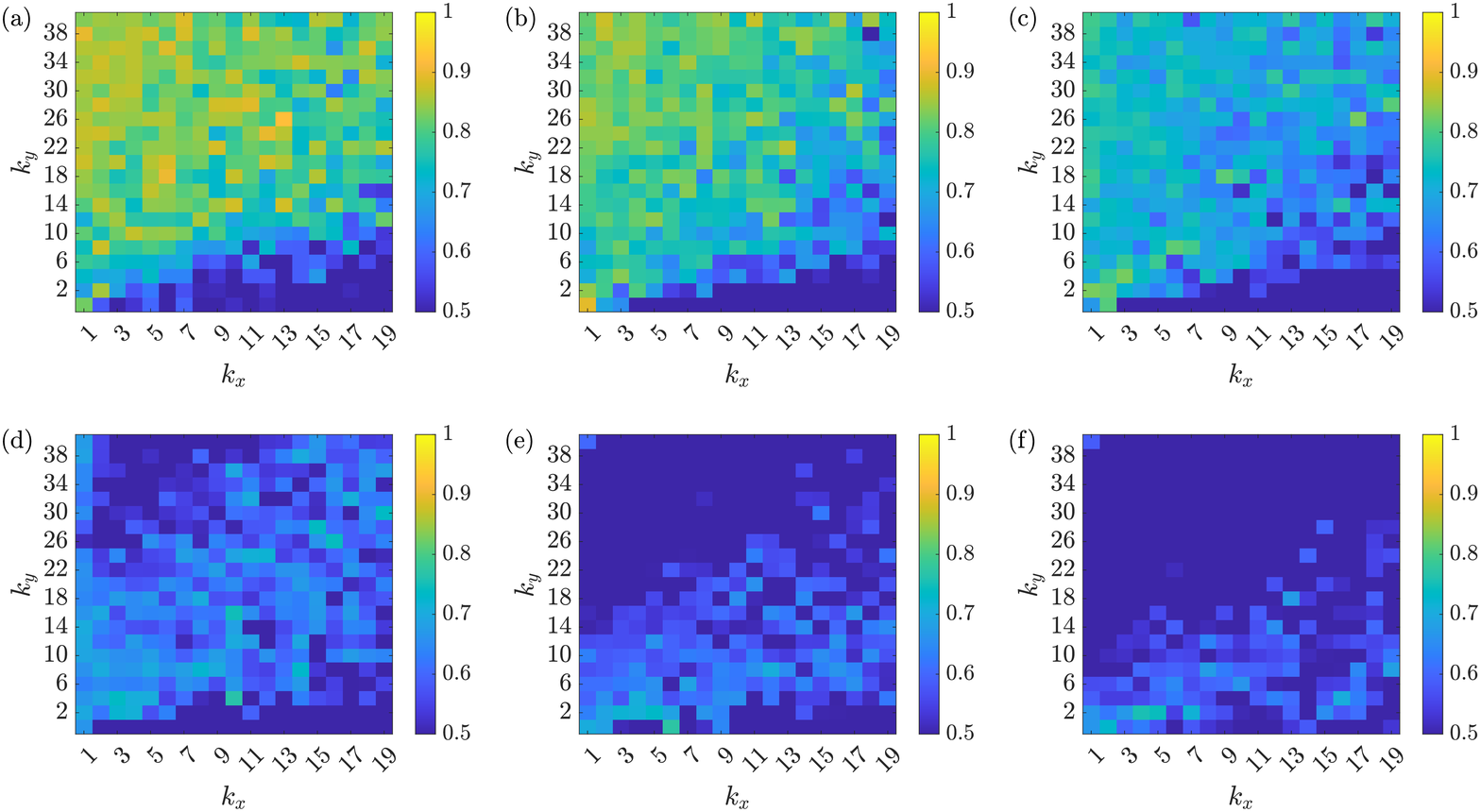}
\caption{Projection of the leading resolvent mode onto the leading SPOD mode for the wave speeds $c^+ = 10.3, ~ 12.0, ~ 13.7, ~ 15.5, ~ 17.2, ~ 18.9$ in ascending order.} \label{fig:molecular projections}
\end{figure}
\begin{figure} 
\centering
\includegraphics[trim=4cm 0.5cm 0cm 1.5cm, clip=true, scale=0.4]{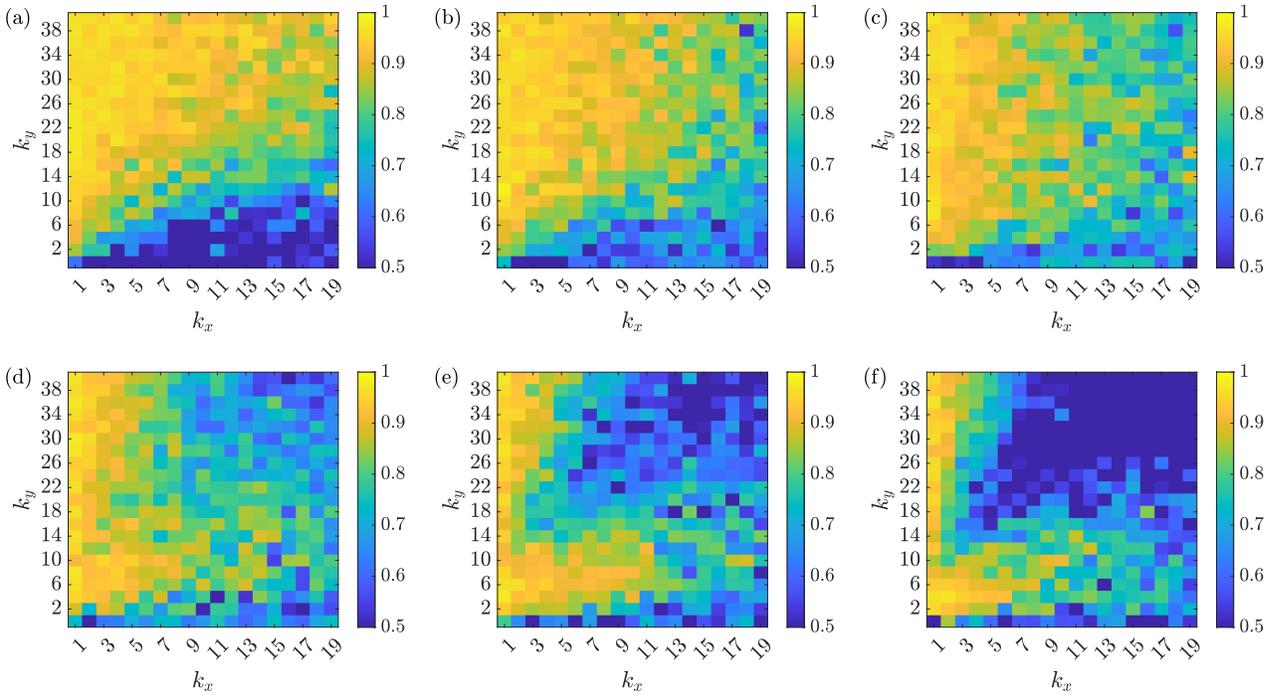}
\caption{Projection of the leading eddy mode onto the leading SPOD mode for the wave speeds $c^+ = 10.3, ~ 12.0, ~ 13.7, ~ 15.5, ~ 17.2, ~ 18.9$ in ascending order.} \label{fig:eddy projections}
\end{figure}

Figure~\ref{fig:molecular projections} illustrates $\gamma$ for wavenumbers that satisfy $1 \leq k_x \leq 19$ and $0 \leq k_y \leq 40$. These wavenumber pairs contain most of the kinetic energy in the flow and include structures associated with the near-wall cycle. Streamwise-constant modes are excluded since the wave speed is undefined for $k_x = 0$. They will be discussed in greater detail in Sec.~\ref{sec:mode shapes}. The colorbar in Fig.~\ref{fig:molecular projections} is restricted to a range of $[0.5,~1]$ to facilitate identification of wavenumber pairs where there is significant overlap between the SPOD and resolvent modes. Each panel in Fig.~\ref{fig:molecular projections} represents a different wave speed. The lowest wave speed considered is $c^+ = 10.3$ in Fig.~\ref{fig:molecular projections}(a) and $c^+$ increases at increments of approximately 1.7, culminating in a wave speed of $c^+ = 18.9$ in Fig.~\ref{fig:molecular projections}(f). The lower bound on $c^+$ is motivated by the near-wall streaks being most energetic at a wall-normal location of $z^+ = 15$ where $U^+ \approx 10$ \cite{McKeon10}. The upper bound on $c^+$ is chosen since it is approximately $c^+ = U^+_{CL} - 2$, which translates to a wall-normal location of $z = -h/2$. Past studies \cite{Morra19, Symon20, Morra21} have noted good agreement between SPOD and eddy modes for structures that are most energetic at this wall-normal location. The eddy viscosity profile $\nu_T(z)$ also reaches a maximum at $z = -h/2$, an observation that is shown to be significant in Sec.~\ref{sec:mode shapes}. 

Figure~\ref{fig:molecular projections} shows that there is good agreement between the leading SPOD and resolvent modes for $c^+ = 10.3$. Despite some outliers, Fig.~\ref{fig:molecular projections}(a) indicates that the highest projections are for modes that satisfy $k_x < k_y$. This is consistent with the results of Ref.~\cite{Abreu20} who noted that the lift-up mechanism leads to large amplification for high aspect ratio scales where $\AR = k_y/k_x$. As the wave speed increases, however, the values of $\gamma$ decline quite significantly, indicating poorer predictions of the mode shapes from resolvent analysis. Figure~\ref{fig:eddy projections} presents the projection coefficient for eddy analysis. It is observed that there is significantly better agreement between the leading SPOD and eddy modes at all wave speeds in comparison to the standard resolvent in Fig.~\ref{fig:molecular projections}. There are, however, some similarities between $\gamma$ and $\gamma^e$. First, the highest values of $\gamma^e$ are obtained for high aspect ratio structures. Second, the wave speed has a major influence on $\gamma^e$. In general, the SPOD and eddy modes overlap less as the wave speed is increased. Unlike $\gamma$, however, there are several wavenumber pairs for which $\gamma^e > 0.9$ at the largest wave speed. The largest $\gamma^e$ are clustered around the largest scales in Fig.~\ref{fig:eddy projections}(f) whereas in Fig.~\ref{fig:eddy projections}(a), they are centered around the near-wall cycle mode $(k_x,k_y) = (4,30)$. 


\subsection{Frequency response} \label{sec:frequency analysis}

The previous sections show that there is a trade-off when eddy viscosity is added to the resolvent operator. On one hand, eddy viscosity distorts the linear dynamics of the operator such that the low-rank map resembles less the turbulent kinetic energy spectrum. On the other hand, the eddy modes have larger projections onto SPOD modes than their resolvent mode counterparts. Although these comparisons have been studied for a variety of wave speeds, the impact of eddy viscosity on the resolvent frequency response has yet to be analyzed. In order to be consistent with previous sections, the frequency response is considered from a wave speed point of view. The impact of eddy viscosity can be better appreciated by considering a single wavenumber pair initially before investigating all energetic wavenumber pairs as done in the previous sections. The wavenumber pair $(k_x,k_y) = (4,30)$, which corresponds to $(\lambda_x^+,\lambda_y^+) = (864,116)$, is selected as it corresponds to the near-wall streaks in the DNS used in this study. 

To gain a better understanding of what the addition of eddy viscosity does to modify the frequency response of the resolvent operator, we look at a range of eddy viscosity based models, where the strength of the eddy viscosity is gradually increased from zero (equivalent to the standard resolvent) to the full eddy-viscosity. The strength of the eddy viscosity is adjusted artificially by introducing the scaling factor $\mathcal{S}$ such that Eq.~(\ref{eq:Cess}) becomes
\begin{equation} \label{eq:S parameter}
\nu_T(z) = \frac{\nu}{2} \left(1 + \mathcal{S} \left[ \frac{\kappa}{3} (1-z^2) (1+2z^2)( 1-\text{exp} (|z-1|Re_{\tau}/A) \right]^2 \right)^{1/2} + \frac{\nu}{2},
\end{equation}
where $\mathcal{S} \in [0,~1]$. Setting $\mathcal{S} = 0$ or $\mathcal{S} = 1$ is equivalent to resolvent analysis or eddy analysis, respectively. A similar parameter was introduced by Ref.~\cite{Gupta21} to derive a scale-dependent eddy viscosity for linear estimation of a turbulent channel flow at $Re_{\tau} = 2003$. 

\begin{figure} 
\centering
\includegraphics[trim=0cm 0cm 0cm 0cm, clip=true, scale=0.4]{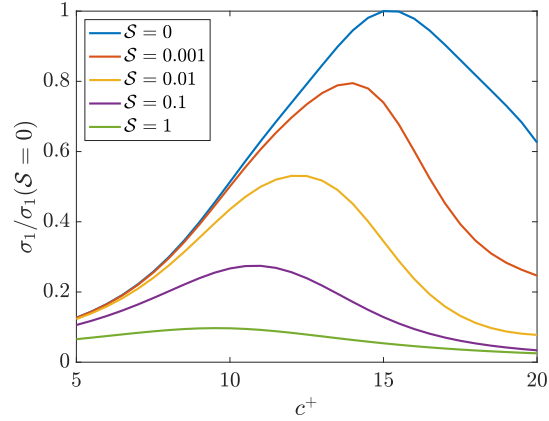}
\caption{Most energetic wave speed for $(k_x,k_y) = (4,30)$ as a function of different strengths of the eddy viscosity as denoted by $\mathcal{S}$.} \label{fig:S}
\end{figure}
The first singular value $\sigma_1$ is plotted against $c^+$ in Fig.~\ref{fig:S} for various strengths of eddy viscosity, i.e. different values of $\mathcal{S}$. As the value of $\mathcal{S}$ increases, both the maximum amplification and the most amplified wave speed decrease. For this particular scale, moreover, the maximum amplification declines by a factor of 10 and the most amplified wave speed slows down substantially from $c^+ = 15.5$ to $c^+ = 9.5$.  A crude explanation for this behavior is that the damping supplied by eddy viscosity results in slower, less amplified structures. Since resolvent analysis has absolutely no damping other than molecular viscosity, the structures are allowed to convect more quickly. It can, therefore, be expected that for an arbitrary scale the most amplified wave speed predicted by resolvent analysis is going to be greater than that predicted by eddy analysis. 

\begin{figure} 
\centering
\includegraphics[trim=4cm 0cm 0cm 0cm, clip=true, scale=0.4]{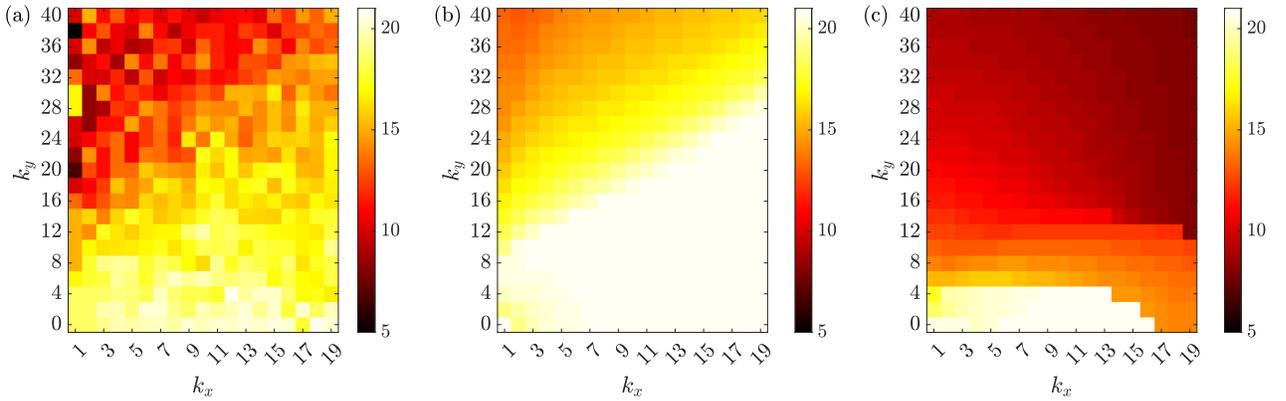}
\caption{Most energetic wave speed computed from (a) DNS compared to the most amplified wave speed predicted by (b) resolvent analysis and (c) eddy analysis.} 
\label{fig:cmax}
\end{figure}

This hypothesis is tested in Fig.~\ref{fig:cmax}, which compares the most energetic wave speed from DNS to the most amplified wave speed predicted by resolvent and eddy analysis. The only similarity among the three panels is that the most energetic/amplified wave speed is primarily governed by the spanwise wavenumber. Wider structures, i.e. those with small spanwise wavenumbers, travel faster than relatively less wide structures. The streamwise wavenumber plays a bigger role in DNS and resolvent analysis than it does for eddy analysis. Figures~\ref{fig:cmax}(a) and (b) show that for fixed $k_y$, the most energetic/amplified wave speed increases as a function of $k_x$. Thus, the trend for $k_x$ is different from $k_y$ in that longer structures, i.e. those with smaller $k_x$, travel slower, while wider structures, i.e. those with smaller $k_y$ travel faster. Another interpretation of these trends is that higher aspect ratio structures convect more slowly and are thus more energetic closer to the wall. Lower aspect ratio structures, meanwhile, convect more quickly and are thus more energetic away from the wall. These trends are consistent with observations from Ref.~\cite{Hwang15} among others who report aspect ratios of approximately $\AR = 8$ for near-wall coherent motions and $\AR = 2-3$ for large-scale outer motions.

\begin{figure} 
\centering
\includegraphics[trim=0cm 0cm 0cm 0cm, clip=true, scale=0.4]{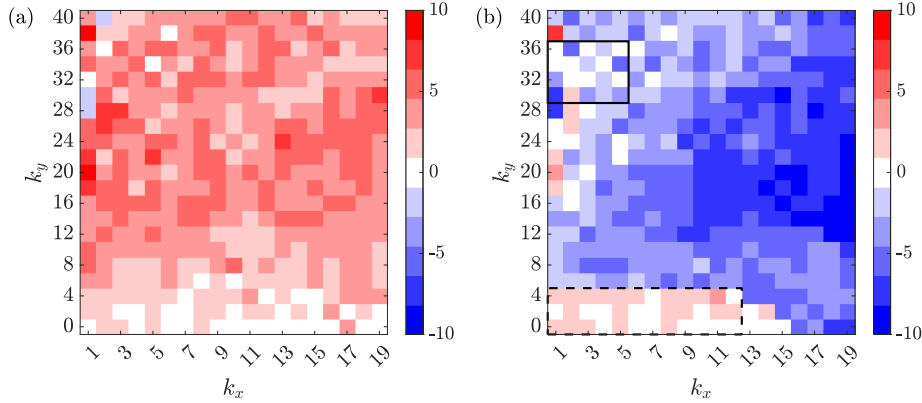}
\caption{Difference between the most amplified wave speed predicted by (a) resolvent and (b) eddy analysis compared to the most energetic wave speed computed in DNS.}
\label{fig:delc}
\end{figure}

Figure~\ref{fig:delc} plots the difference between the most amplified wave speed predicted by resolvent/eddy analysis and the most energetic wave speed computed from DNS. Red squares indicate that the most amplified wave speed is too high and blue squares that the most amplified wave speed is too low. It is striking how large the differences are particularly for the relatively small Reynolds number considered in this study which restricts the range of wave speeds that can be expected in the flow. Figure~\ref{fig:delc}(a) shows that the most amplified structures in resolvent analysis travel faster than their true speeds whereas in Fig.~\ref{fig:delc}(b), the most amplified structures in eddy analysis travel slower. Despite the large differences between DNS and the linear analyses, there are two regions in Fig.~\ref{fig:delc}(b) where the discrepancy is small. The first is denoted by the dashed, black rectangle and corresponds to small values of both $k_x$ and $k_y$, where $c^+ \approx 18.9$. The second is denoted by the solid black rectangle, which contains wavenumber pairs in the range $(k_x,k_y) = (1 \leq  k_x \leq 5, 30 \leq k_y \leq 34)$ where $c^+ \approx 10$. As will be discussed in the next section, the reasons for why these two wavenumber regions are estimated reasonably well by the eddy model can be related back to the wall-normal profile of the eddy viscosity $\nu_T(z)$. 

\section{Mode shapes and the Cess eddy viscosity profile} \label{sec:mode shapes}

The previous section analyzed the linear predictions of resolvent and eddy analysis using scalar quantities such as the low-rank maps, projection coefficients, and most amplified wave speeds. This section considers the wall-normal profiles of both mode shapes in Sec.~\ref{sec:mode structures} and the Cess eddy viscosity profile itself in Sec.~\ref{sec:Cess scaling} to predict the agreement between SPOD and resolvent/eddy modes at higher Reynolds numbers. 

\subsection{SPOD, resolvent, and eddy modes} \label{sec:mode structures}

\begin{table}
	\begin{center}
		\def~{\hphantom{0}}
		\begin{tabular}{cccccccccc}
			Mode & $k_x$ & $k_y $ & $c^+_{\text{max,SPOD}} $ & $c^+_{\text{max,resolvent}} $ & $c^+_{\text{max,eddy}} $ & $\gamma $ &  $\gamma^e$  & $\gamma_{95}$ & $\gamma_{95}^e$ \\
			0 & 0 & 4 & $\omega = 0$ & $\omega = 0$ & $\omega = 0$ & 0.898 & 0.978 & 6 & 2 \\
			1 & 1 & 2 & 18.5 & 19.0 & 18.5 & 0.663 & 0.956 & 36 & 2 \\
			2 & 2 & 8 & 16.8 &  20.5 & 13.8  &  0.586 & 0.907 & 56 & 8 \\
			3 & 4 & 30 & 10.1 & 15.3 & 9.50 & 0.864 & 0.959 & 40 & 2\\
		\end{tabular}
		\caption{The wavenumber triplets for modes 0-3 along with predictions from resolvent and eddy analysis.}
		\label{tab:triplets}
	\end{center}
\end{table}

The SPOD, resolvent, and eddy mode shapes are compared for four wavenumber triplets that are described in Table~\ref{tab:triplets}. Mode 0 is a streamwise-constant mode and its most energetic frequency is $\omega = 0$. The other wavenumber pairs are chosen such that Modes 1, 2, and 3 have their peak energy around $z/h = -h/2$, $z/h = -h/4$, and $z^+ = 15$, respectively. The most energetic wave speed from SPOD is reported as $c_{\text{max,SPOD}}^+$. Table~\ref{tab:triplets} reports $c^+_{\text{max,resolvent}}$ and $c^+_{\text{max,eddy}}$, the most amplified wave speeds identified by resolvent and eddy analysis, respectively. The wave speed selected for plotting purposes, however, is kept fixed at $c_{\text{max,SPOD}}^+$. The purpose of comparing $c_{\text{max,SPOD}}^+$ and $c^+_{\text{max,eddy}} $ is to emphasize that eddy analysis produces good predictions when $c_{\text{max,SPOD}}^+ \approx c^+_{\text{max,eddy}} $. The parameter $c_{\text{max,resolvent}}^+$, meanwhile, does not provide information about the quality of predictions from resolvent analysis. Finally, Table~\ref{tab:triplets} quantifies the number of resolvent/eddy modes needed such that the velocity from DNS is reconstructed to $95\%$ accuracy. In other words, the number of modes required to obtain a value of $\gamma=0.95$ using Eq.~\ref{eq:projection gamma}. 

\begin{figure} 
\centering
\includegraphics[trim=4cm 0cm 0cm 0cm, clip=true, scale=0.4]{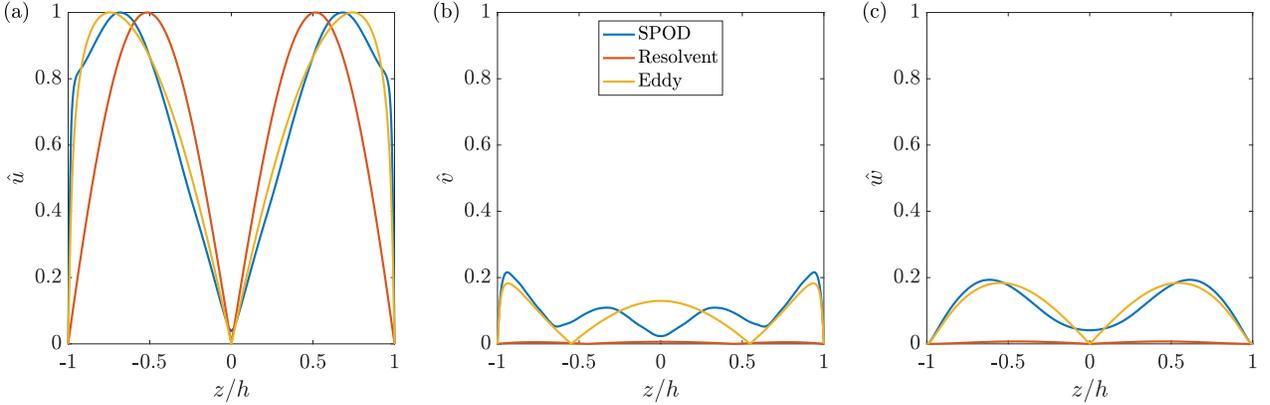}
\caption{The (a) streamwise, (b) spanwise, and (c) wall-normal component of the leading SPOD, resolvent, and eddy mode for Mode 0 for which $\boldsymbol{k} = (0,4,\omega=0)$.} \label{fig:mode0}
\end{figure}

Figure~\ref{fig:mode0} presents the SPOD, resolvent, and eddy mode for Mode 0. As indicated in Table~\ref{tab:triplets}, both resolvent and eddy analysis correctly identify $\omega = 0$ as the most amplified frequency. The projection coefficients $\gamma$ and $\gamma^e$ are also large for this choice of $\boldsymbol{k}$. The number of resolvent or eddy modes needed to achieve $\gamma = 0.95$, therefore, is $\gamma_{95} = 6$ and $\gamma^e_{95} = 2$, respectively. The agreement between the resolvent and SPOD modes themselves, however, is not as compelling as $\gamma = 0.898$ might suggest. Although the streamwise velocity component is predicted reasonably well, the spanwise and wall-normal components are significantly underestimated due to the high non-normality of the resolvent operator \cite{Trefethen93, Schmid01}. The eddy operator is also non-normal but the addition of eddy viscosity results in a more normal operator \citep{Symon21}. Since the streamwise component is also dominant for the SPOD mode, it has a disproportionate influence on $\gamma$ thus resulting in a higher value than might be expected from visual inspection of the mode shapes. The agreement between the SPOD and eddy modes, on the other hand, is good for all velocity components, resulting in $\gamma^e = 0.978$. 

\begin{figure} 
\centering
\includegraphics[trim=4cm 0cm 0cm 0cm, clip=true, scale=0.4]{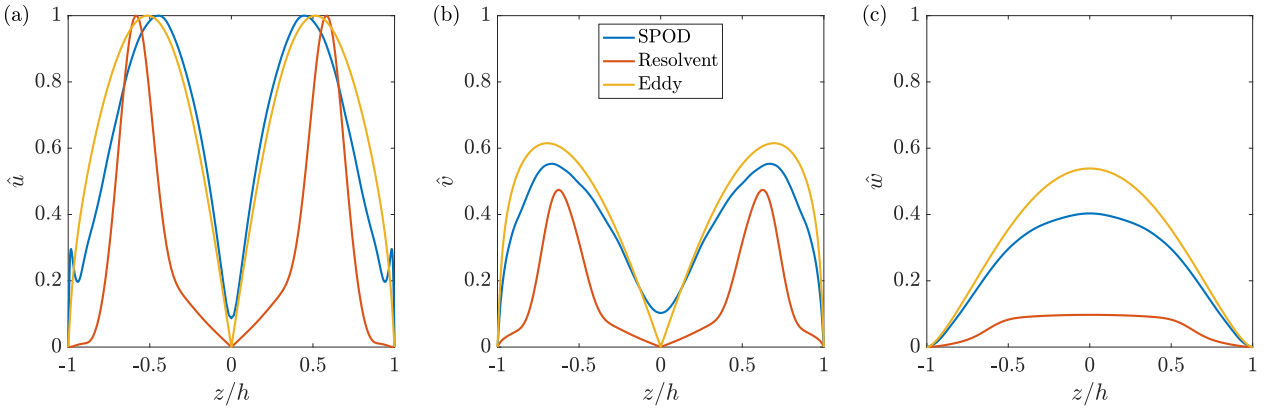}
\caption{The (a) streamwise, (b) spanwise, and (c) wall-normal component of the leading SPOD, resolvent, and eddy mode for Mode 1 for which $\boldsymbol{k} = (1,2,18.5)$.} \label{fig:mode1}
\end{figure}

Mode 1 is an energetic structure in the outer region of the flow with a wavenumber triplet of $\boldsymbol{k} = (1,2,18.5)$. Table~\ref{tab:triplets} shows that $c^+_{\text{max,SPOD}} = 18.5$ and this is in good agreement with predictions from resolvent and eddy analysis. Figure~\ref{fig:mode1} illustrates that while the resolvent modes do not agree well with the SPOD modes, the eddy modes show reasonable agreement with SPOD. This is reflected in the projection coefficients as $\gamma^e = 0.956$ is higher than $\gamma = 0.663$. It also takes significantly fewer eddy modes to reconstruct the leading SPOD mode since as $\gamma^e_{95} = 2$ whereas $\gamma_{95} = 36$. Refs.~\cite{Rosenberg19,Morra19,Symon21} also observed that, when using resolvent modes, many suboptimal modes are needed to reconstruct the velocity field for high aspect ratio structures. One reason is that the streamwise and spanwise components of velocity are highly localized in the wall-normal direction due to the critical-layer mechanism \cite{McKeon10,Morra19,Madhusudanan19,Vadarevu19,Symon20}. The eddy modes, on the other hand, are smoothed out in $z$ by the eddy viscosity. Another factor is that, similar to Mode 0, the wall-normal velocity component is underestimated by resolvent analysis as seen in Fig.~\ref{fig:mode1}(c). The agreement between the SPOD and eddy modes, meanwhile, is very good for all velocity components and wall-normal locations except the near-wall region of the streamwise component, as seen in Fig.~\ref{fig:mode1}(a).

\begin{figure} 
\centering
\includegraphics[trim=4cm 0cm 0cm 0cm, clip=true, scale=0.4]{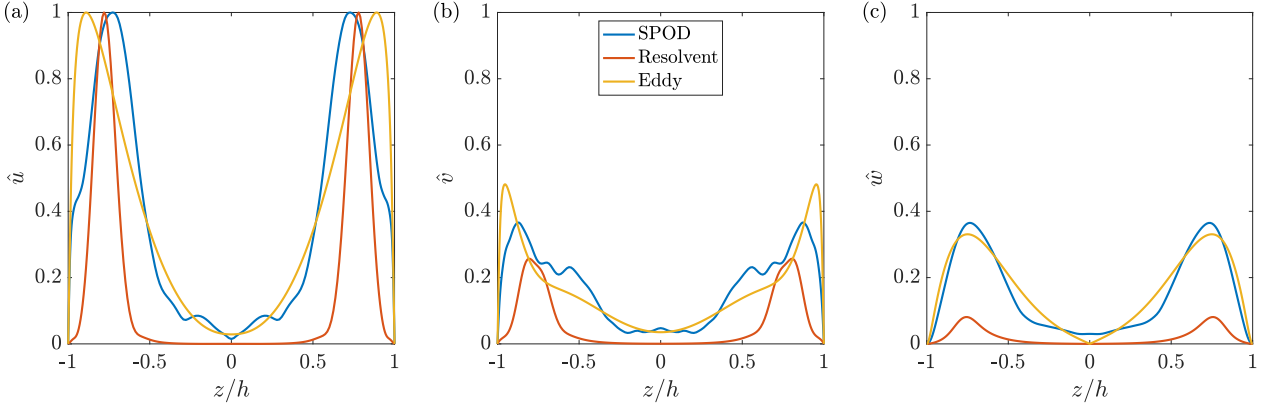}
\caption{The (a) streamwise, (b) spanwise, and (c) wall-normal component of the leading SPOD, resolvent, and eddy mode for Mode 2 for which $\boldsymbol{k} = (2,8,16.8)$.} \label{fig:mode2}
\end{figure}

Mode 2 is another structure in the outer region of the flow with a wavenumber triplet of $\boldsymbol{k} = (2,8,16.8)$. The streamwise velocity component is most energetic at $z/h = -0.75$ which is closer to the wall in comparison to $z/h = -0.5$ for Mode 1. Neither resolvent nor eddy analysis are capable of predicting the most energetic wave speed for Mode 2. Consistent with the trends observed in Fig.~\ref{fig:delc}, resolvent analysis predicts a wave speed that is too fast while eddy analysis predicts a wave speed that is too slow.  As explained earlier, for plotting the resolvent and eddy modes in Fig.~\ref{fig:mode2}, $c^+_{resolvent}$ and $c^+_{eddy}$ are chosen to be the maximum wave-speed as identified by SPOD, i.e. $c^+_{max,SPOD}$. It can be remarked that the projection coefficients belie the true agreement between various mode shapes. When considering the resolvent modes, although $\gamma = 0.586$ seems low, the modes still capture the wall-normal location of the peak streamwise energy reasonably well. However, similar to Mode 1, the modes are localized about the critical-layer and the wall-normal component is underestimated resulting in a relatively low $\gamma$. The eddy modes appear to be in significantly better agreement as $\gamma^e = 0.907$. Unlike resolvent analysis, however, the wall-normal location of the peak in streamwise and spanwise energies of the eddy modes fall below their true locations, a trend that becomes more apparent for higher Reynolds numbers (see Ref.~\cite{Symon20}). Even though it is not explicitly shown here for the sake of brevity, for structures with $c^+_{\text{max,eddy}} < c^+_{\text{max,SPOD}}$, the peak streamwise and spanwise energies of the eddy mode are located below the correct wall-normal location identified by SPOD. Mode 2, consequently, is the only wavenumber triplet for which $\gamma^e_{95} = 8$ since suboptimal modes are required to ``lift'' the structure to its proper height. 58 modes, meanwhile, are required for resolvent modes to achieve $\gamma = 0.95$. 

\begin{figure} 
\centering
\includegraphics[trim=4cm 0cm 0cm 0cm, clip=true, scale=0.4]{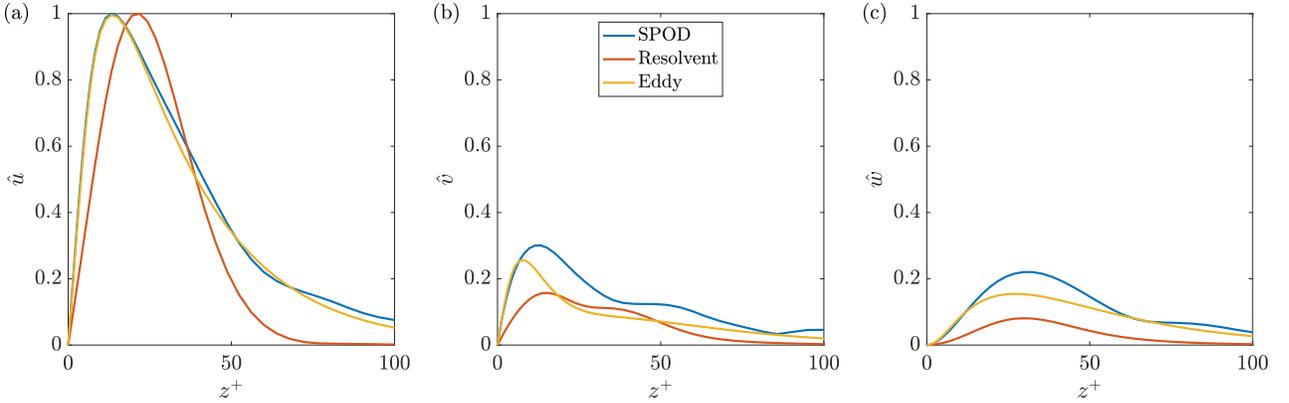}
\caption{The (a) streamwise, (b) spanwise, and (c) wall-normal component of the leading SPOD, resolvent, and eddy mode for Mode 3 for which $\boldsymbol{k} = (4,30,10.1)$.} \label{fig:mode3}
\end{figure}

Figure~\ref{fig:mode3} considers Mode 3, which is representative of the near-wall cycle. The agreement between the SPOD and resolvent wave speeds is poor but agreement between the mode shapes is high as $\gamma = 0.864$. Similar to Mode 0, the streamwise velocity component is the dominant component for Mode 3, thus relegating the influence of the other two velocity components in computing $\gamma$. As seen in Table~\ref{tab:triplets}, eddy analysis is close to identifying the most energetic wave speed. Furthermore, the SPOD and eddy mode shapes in Fig.~\ref{fig:mode3}, are in excellent agreement for the streamwise velocity component and good agreement for the spanwise and wall-normal velocity components, resulting in $\gamma^e = 0.959$. It is clear, furthermore, that eddy modes are a more efficient basis since $\gamma^e_{95} = 2$ compared to $\gamma_{95} = 40$. Similar agreement between DNS and eddy analysis predictions for near-wall structures has also been reported by Refs.~\cite{Morra19, Morra21}. 

\subsection{Cess eddy viscosity and effective Reynolds number} \label{sec:Cess scaling}

\begin{figure} 
\centering
\includegraphics[trim=0cm 0cm 0cm 0cm, clip=true, scale=0.4]{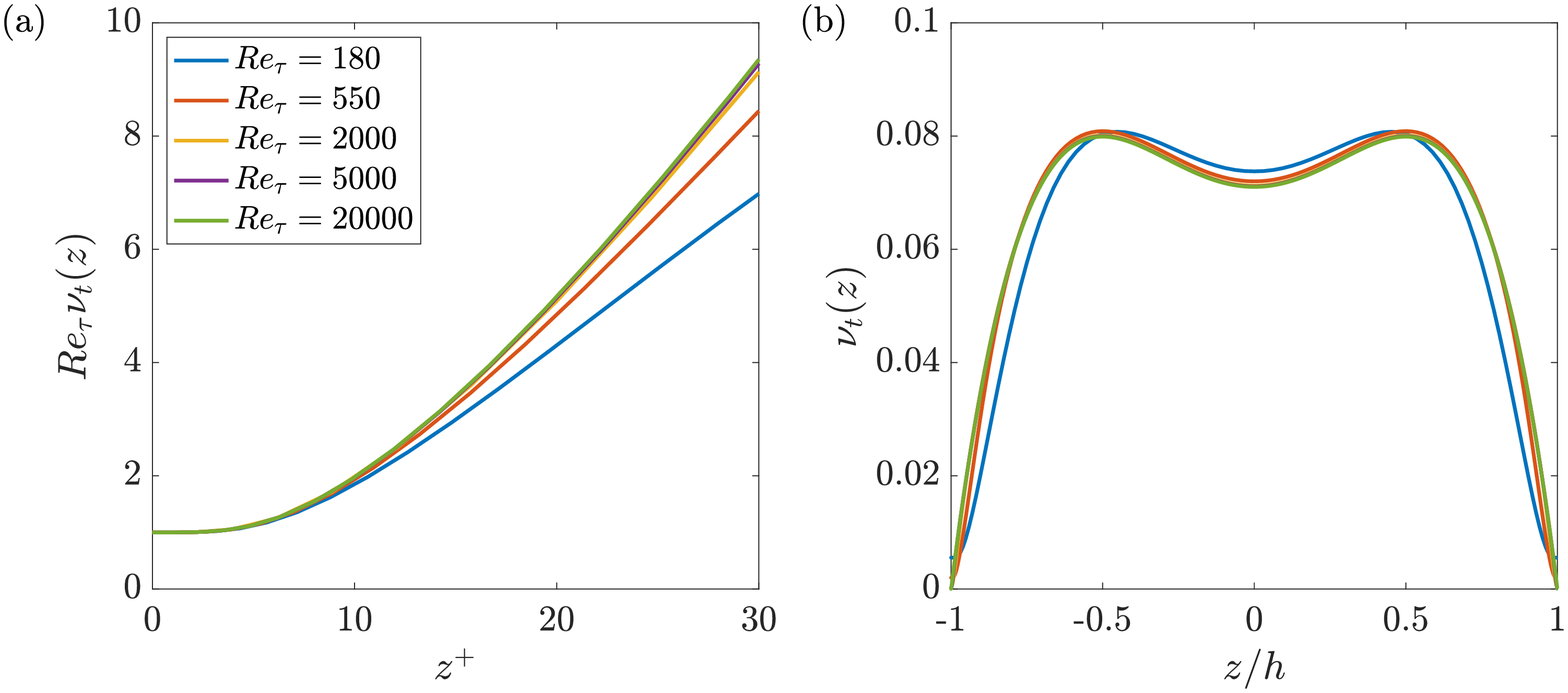}
\caption{Wall-normal profiles of (a) $Re_{\tau}\nu_T$ and (b) $\nu_T$ for various $Re_\tau$.} \label{fig:eddyRe}
\end{figure}

In this section, the Cess eddy viscosity profile is studied in greater detail in order to predict the agreement between SPOD and resolvent/eddy modes at higher Reynolds numbers. The profiles are obtained for a range of Reynolds numbers using Eq.~(\ref{eq:Cess}). It is particularly interesting to look at the profiles around $z^+ = 15$ and $z = -h/2$ due to the good agreement between SPOD and eddy analysis for Modes 1 and 3 that have their peak energies at these wall-heights. In Fig.~\ref{fig:eddyRe}(a), the eddy viscosity profiles are pre-multiplied by $Re_{\tau}$, resulting in a collapse near the wall. The profiles for $Re_{\tau} = 180$ and $Re_{\tau} = 550$ begin to diverge from the other Reynolds numbers around $z^+ = 10$  and $z^+ = 20$, respectively, but the agreement among all profiles is good, particularly around $z^+ = 15$. This wall-normal location is important since it coincides with the location of the structures associated with the near-wall cycle that travel at a wave speed of $c^+ = 10$.  It therefore seems reasonable to assume that, for all Reynolds numbers, the eddy analysis will correctly predict the structures that are located at these wall-heights of $z^+\approx15$ and therefore convect at wave speeds around $c^+ = 10$. Consistent with this observation, Ref.~\cite{Morra19} observed that at $Re_{\tau} = 1007$, the eddy analysis is able to get reasonable predictions for the structure $(k_x,k_y)=(14,63)$ convecting at $c^+=10$. 

\begin{figure} 
\centering
\includegraphics[trim=4cm 0cm 0cm 0cm, clip=true, scale=0.4]{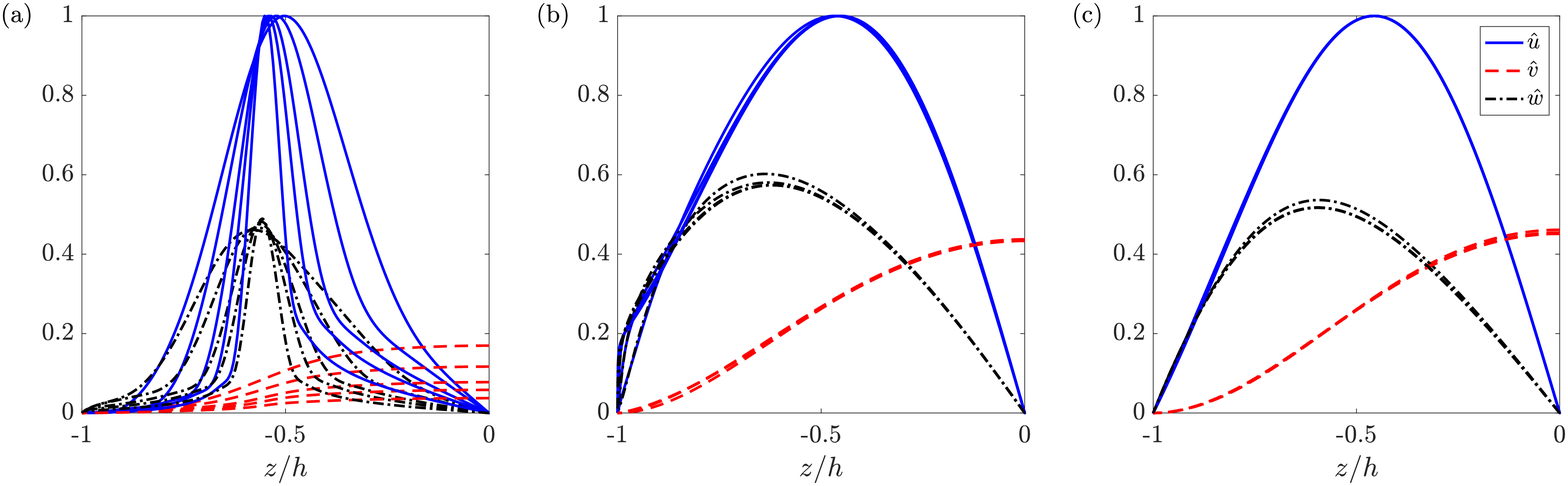}
\caption{Mode shapes for $\boldsymbol{k} = (1,2,U^+_{CL}-2)$ from (a) resolvent analysis, (b) eddy analysis, and (c) eddy analysis setting $Re_T = 12.5$. The Reynolds numbers range from $Re_{\tau} = 180$ up to $Re_{\tau} = 20000$.} \label{fig:constantRe}
\end{figure}

In Fig.~\ref{fig:eddyRe}(b), $\nu_T(z)$ is plotted in outer units for Reynolds numbers in the range $180 \leq Re_{\tau} \leq 20000$. The profiles for all Reynolds numbers other than $Re_{\tau} = 180$ are virtually indistinguishable in the outer region. The largest differences occur near and at the walls where $\nu_T(0) = \nu_T (2h) = \nu$. The maximum value of $\nu_T$ is  0.08 and occurs at $z = \pm h/2$. Mode 1, for which the eddy analysis gives good predictions, was also found to be most energetic at this wall-height of $z = \pm h/2$. Since $\nu_T$ at $z = \pm h/2$ remains roughly constant with Reynolds number, it can be hypothesized that the eddy mode shapes for wavenumbers with $c^+_{\text{max}} = U^+(\pm h/2) \approx U^+_{CL} - 2$ are unaffected by $Re_{\tau}$. This hypothesis is tested in Fig.~\ref{fig:constantRe} where the mode shapes of structures with $c^+_{\text{max}} \approx U^+_{CL} - 2$ are compared across different $Re_\tau$. A second hypothesis, which is also tested in Fig.~\ref{fig:constantRe}, is that the wall-normal-varying eddy viscosity $\nu_T(z)$ can be replaced by a constant eddy viscosity model, i.e. $\nu_T = Re_T$, where $Re_T$ is the effective Reynolds number. For the case of a turbulent jet, References~\cite{Pickering21,Kuhn21,Kuhn22} showed that a constant eddy viscosity model can improve the agreement between SPOD and eddy modes.

Figure~\ref{fig:constantRe} compares the mode shapes for $\boldsymbol{k} = (1,2,U^+_{CL}-2)$ from resolvent analysis, eddy analysis, and eddy analysis setting $Re_T = 1/(\text{max}(\nu_T)) = 12.5$ as done in \cite{Hwang16}. If $\nu_T = Re_T$, then only the mean profile is affected by changes in Reynolds number. All sets of modes are computed for the same Reynolds numbers that appeared in Fig.~\ref{fig:eddyRe} and are normalized by the maximum value of the streamwise velocity component. The resolvent modes in Fig.~\ref{fig:constantRe}(a) are influenced heavily by the choice of $Re_{\tau}$. The streamwise and spanwise velocity components become increasingly localized about the critical layer at $z = -h/2$ as the Reynolds number increases. The shape of the wall-normal component, on the other hand, is roughly constant but its magnitude relative to the wall-parallel velocity components decreases with increasing Reynolds number. 

The impact of Reynolds number on the eddy and constant $Re_T$ modes in Figs.~\ref{fig:constantRe}(b,c) is negligible. The only difference among the eddy modes in Fig.~\ref{fig:constantRe}(b) is that for $Re_{\tau} = 180$, the streamwise and spanwise velocity components are less attached to the wall in comparison to the other profiles which appear more blunt, i.e. flatter, near the wall. A similar difference emerges between the eddy and constant $Re_T$ modes in that neither the streamwise nor spanwise components exhibit blunt behavior near the wall for any Reynolds number considered. The agreement between eddy and constant $Re_T$ modes, nonetheless, is remarkable given that the constant eddy viscosity model is so simple. 

\begin{figure} 
\centering
\includegraphics[trim=4cm 0cm 0cm 0cm, clip=true, scale=0.4]{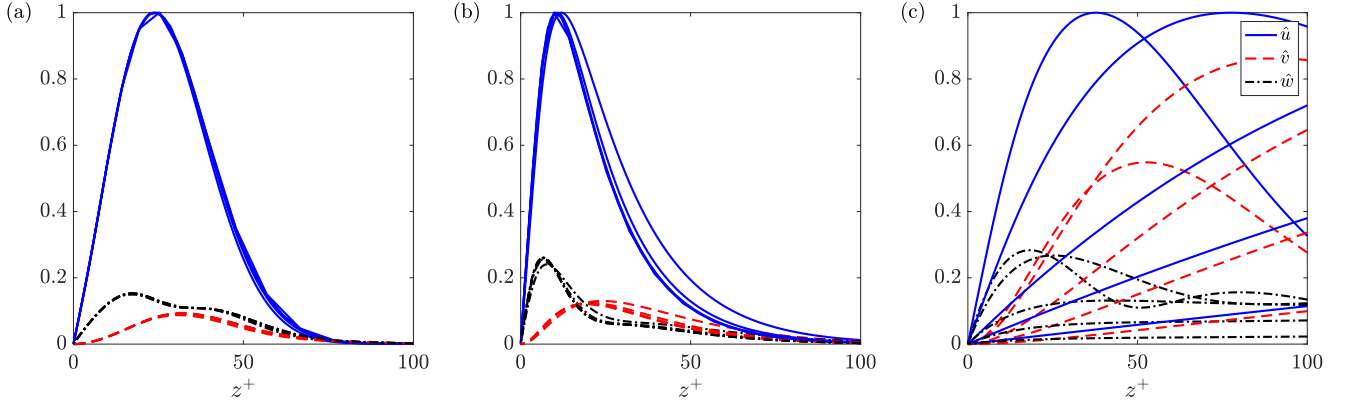}
\caption{Mode shapes for $(k_x^+,k_y^+,c^+) = (2\pi/1000,2\pi/100,10)$ from (a) resolvent analysis, (b) eddy analysis, and (c) eddy analysis setting $Re_T = 12.5$. The Reynolds numbers range from $Re_{\tau} = 180$ up to $Re_{\tau} = 20000$.} \label{fig:inner}
\end{figure}

The applicability of the constant eddy viscosity model, however, is limited to scales that have maximum streamwise energy at $z/h = \pm0.5$. Figure~\ref{fig:inner} compares the mode shapes for $(k_x^+,k_y^+,c^+) = (2\pi/1000,2\pi/100,10)$, which are representative of the near-wall cycle, from resolvent analysis, eddy analysis, and eddy analysis setting $Re_T = 1/(\text{max}(\nu_T)) = 12.5$. The Reynolds number has little impact on the resolvent and eddy modes but has a major impact on the constant $Re_T$ modes. As the Reynolds number increases, the mode shapes become increasingly less localized about the critical layer. The location of the peak energy of all three velocity components, furthermore, gradually shifts closer to the channel centerline and away from the wall. Although it is not shown in the interest of brevity, the same trends can be observed for other values of $Re_T$. The wall-varying profile of the eddy viscosity near the wall, consequently, is essential to capture the correct mode shapes for wavenumber triplets that are associated with the near-wall cycle. 

\section{Nonlinear Energy transfer and eddy viscosity} \label{sec:energy}

The improved agreement between SPOD and eddy modes indicates that eddy viscosity is able to partially model the effect of $\hat{\boldsymbol{f}}$ for particular scales such as Modes 1 and 3. In this section, the energy transfers that are introduced by the eddy viscosity, herein referred to as eddy dissipation, are examined in greater detail to explain the success or failure of eddy analysis in predicting the correct structures. Section~\ref{sec:eddy dissipation} introduces eddy dissipation which consists of two terms. The first originates from the wall-normal-varying effective Reynolds number $\nu_T(z)-\nu$ and the second from the eddy viscosity gradient $\nu_T' = d \nu_T/dz$. In Sec.~\ref{sec:positive and negative}, the wall-normal profiles of these two transfers are examined in order to determine how they affect the eddy mode shapes. It is demonstrated in Sec.~\ref{sec:DnuT} that artificially adjusting the eddy viscosity gradient can significantly manipulate the eddy mode shapes, particularly in the near-wall region. Section~\ref{sec:discussion} discusses the types of interactions that are modelled by the eddy viscosity.

\subsection{Eddy Dissipation} \label{sec:eddy dissipation}

As explained in Ref.~
\cite{Symon21}, the addition of eddy viscosity introduces new dissipation terms into the kinetic energy balance for each scale. These terms can be derived by expanding the viscous term in Eq.~(\ref{eq:triple}) and Fourier-transforming in the homogeneous directions
\begin{equation} \label{eq:viscous index}
\boldsymbol{\nabla} \cdot \left[\nu_T(\boldsymbol{\nabla} \hat{\boldsymbol{u}} + \boldsymbol{\nabla} \hat{\boldsymbol{u}}^T) \right] = \nu_T \boldsymbol{\nabla}^2 \hat{\boldsymbol{u}} + (\boldsymbol{\nabla} \cdot \nu_T) (\boldsymbol{\nabla} \hat{\boldsymbol{u}} + \boldsymbol{\nabla}\hat{\boldsymbol{u}}^T).
\end{equation}
Rewriting Eq.~(\ref{eq:viscous index}) in index notation, taking the inner product with respect to $u_i$, and averaging the final expression in time yields 
\begin{equation} \label{eq:DVG}
\hat{\mathcal{D}}_T(k_x,k_y,z) = 
\underbrace{- \nu \overline{ \frac{\partial \hat{u}_i}{\partial x_j} \frac{\partial \hat{u}_i}{\partial x_j} }}_{\hat{D}(k_x,k_y,z)}  
\underbrace{- (\nu_T(z)-\nu) \overline{  \frac{\partial \hat{u}_i}{\partial x_j} \frac{\partial \hat{u}_i}{\partial x_j} }}_{\hat{V}(k_x,k_y,z)} 
\underbrace{ + \frac{d\nu_T(z)}{dz} \overline{ \hat{u}_i \left(\frac{\partial \hat{u}_i}{\partial z} + \frac{\partial \hat{w}}{\partial x_i}\right) }}_{\hat{G}(k_x,k_y,z)},
\end{equation}
where $i,j = 1,2,3$ and $\hat{\mathcal{D}}_T$ is dissipation due to molecular and eddy viscosity. $\hat{\mathcal{D}}_T$ can be split into three separate terms: $\hat{D}$ is dissipation due to molecular viscosity, $\hat{V}$ is additional dissipation introduced by a wall-varying effective Reynolds number $\nu_T(z)-\nu$, and $\hat{G}$ is dissipation due to the wall-normal gradient of the eddy viscosity profile. The combined effect of $\hat{V} + \hat{G}$ can be referred to as eddy dissipation, i.e. $\widehat{Edd} = \hat{V} + \hat{G}$. It can be noted that without eddy viscosity, $\widehat{Edd} = \hat{V} = \hat{G} = 0$ . 

\subsection{Positive and negative energy transfers} \label{sec:positive and negative}

\begin{figure} 
\centering
\includegraphics[trim=0cm 0cm 0cm 0cm, clip=true, scale=0.4]{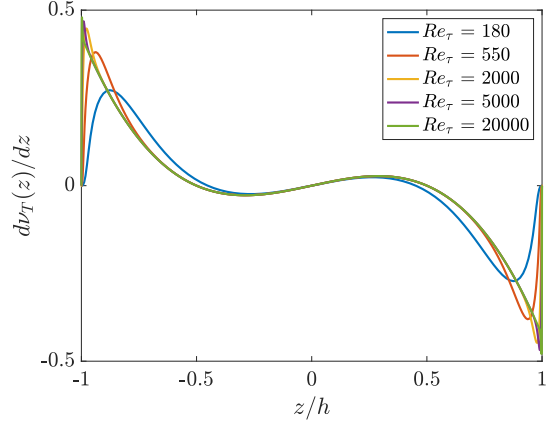}
\caption{Wall-normal profiles of the eddy viscosity gradient $\nu_T'$ for various $Re_\tau$} \label{fig:dnudz}
\end{figure}

It is guaranteed that $\hat{D}$ and $\hat{V}$ are real and negative at all wall-normal locations, but this is not the case for $\hat{G}$. In fact, it can be shown that $\hat{G}$ is likely to be positive near the wall. If the final term of Eq.~(\ref{eq:DVG}) is expanded in full, then $\hat{G}$ becomes
\begin{equation} \label{eq:Gall}
\hat{G} = \frac{d \nu_T}{dz} \overline{\hat{u}\frac{\partial \hat{u}}{\partial z} } + 
\frac{d \nu_T}{dz} \overline{ \hat{v}\frac{\partial \hat{v}}{\partial z} }  + 
2 \frac{d \nu_T}{dz} \overline{ \hat{w} \frac{\partial \hat{w}}{\partial z} }  + 
\frac{d \nu_T}{dz} \overline{ \hat{u} \frac{\partial \hat{w}}{\partial x} }  + 
\frac{d \nu_T}{dz} \overline{ \hat{v} \frac{\partial \hat{w}}{\partial y} }. 
\end{equation}
Assuming that the streamwise velocity component is significantly stronger than the spanwise and wall-normal components, i.e. $\hat{u} \gg \hat{v}, \hat{w}$, and that wall-normal gradients dominate over streamwise and spanwise gradients, Eq.~(\ref{eq:Gall}) can be approximated as
\begin{equation} \label{eq:G1term}
\hat{G} \approx \frac{d \nu_T}{dz} \overline{  \hat{u}\frac{\partial \hat{u}}{\partial z} }. 
\end{equation}
For the lower wall, $\hat{u} = 0$ due to the no-slip condition and both $\hat{u}$ and $\frac{\partial \hat{u}}{\partial z}$ must be either both positive or negative moving away from the wall. In either case, the product of $\hat{u}$ and $\frac{\partial \hat{u}}{\partial z}$ is positive. As shown in Fig.~\ref{fig:dnudz}, $\frac{d \nu_T}{dz} > 0$ near the lower wall so the product of all three terms that appear in Eq.~\ref{eq:G1term} is positive in the near-wall region. Although several approximations have been made to arrive at this result, it is significant because $\hat{V}$ and $\hat{G}$ attempt to model nonlinear transfer processes. If $\hat{G}$ can be positive at a given wall-normal location, then it implies that eddy viscosity can model locally both positive and negative energy transfer. The net energy transfer due to eddy viscosity that is obtained by integrating $\widehat{Edd}$ over the wall-normal direction tends to be negative \cite{Symon21}. 

Before examining profiles of $\hat{V}(z)$ and $\hat{G}(z)$ for specific scales, it is worth commenting further on the eddy viscosity gradient profiles in Fig.~\ref{fig:dnudz}. Similar to the $\nu_T$ profiles in Fig.~\ref{fig:eddyRe}, it can be observed that $Re_\tau$ has a negligible impact on $\nu_T'$ for most wall-normal locations. The maxima and minima of profiles become more extreme and closer to the wall as $Re_\tau$ increases. This localization suggests that $\hat{G}(z)$ is likely to be more concentrated in the near-wall regions and have a larger magnitude for higher Reynolds numbers. 

\begin{figure} 
\centering
\includegraphics[trim=4cm 0cm 0cm 0cm, clip=true, scale=0.4]{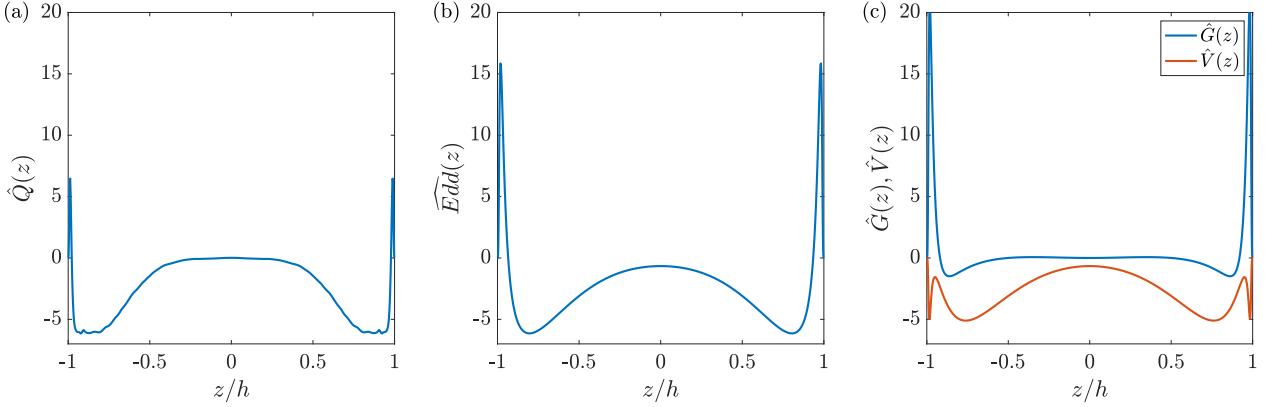}
\caption{(a) Time-averaged nonlinear transfer from DNS for $(k_x,k_y) = (0,4)$ compared to (b) eddy dissipation and (c) its components for $\omega = 0$.} \label{fig:edd streak}
\end{figure}

Figure~\ref{fig:edd streak}(a) compares $\hat{Q}(z) = -\overline{\hat{u}_i\frac{\partial}{\partial x_j}\widehat{u_iu_j}}$, the time-averaged nonlinear transfer from DNS, for $(k_x,k_y) = (0,4)$ with $\widehat{Edd}(z)$ in Fig.~\ref{fig:edd streak}(b). Since the DNS results are averaged in time, they include contributions from all frequencies (wave speeds) whereas the eddy predictions are for the most energetic frequency only. This is a reasonable approximation since $\omega = 0$ dominates over all other frequencies (recall that wave speed is ill-defined for $k_x = 0$ modes). There is good agreement between $\hat{Q}(z)$ and $\widehat{Edd}(z)$ at nearly all wall-normal locations other than the near-wall region. Eddy dissipation correctly predicts positive energy transfer near the wall but its magnitude is too large. In Fig.~\ref{fig:edd streak}(c), $\widehat{Edd}(z)$ is split into the contributions from $\hat{V}(z)$ and $\hat{G}(z)$. Fig.~\ref{fig:edd streak}(c) reinforces that $\hat{G}(z)$ is responsible for the positive energy transfer, as predicted by Eq.~(\ref{eq:G1term}), and that $\hat{V}(z)$ accounts for the bulk of the negative energy transfer. 

\begin{figure} 
\centering
\includegraphics[trim=4cm 0cm 0cm 0cm, clip=true, scale=0.4]{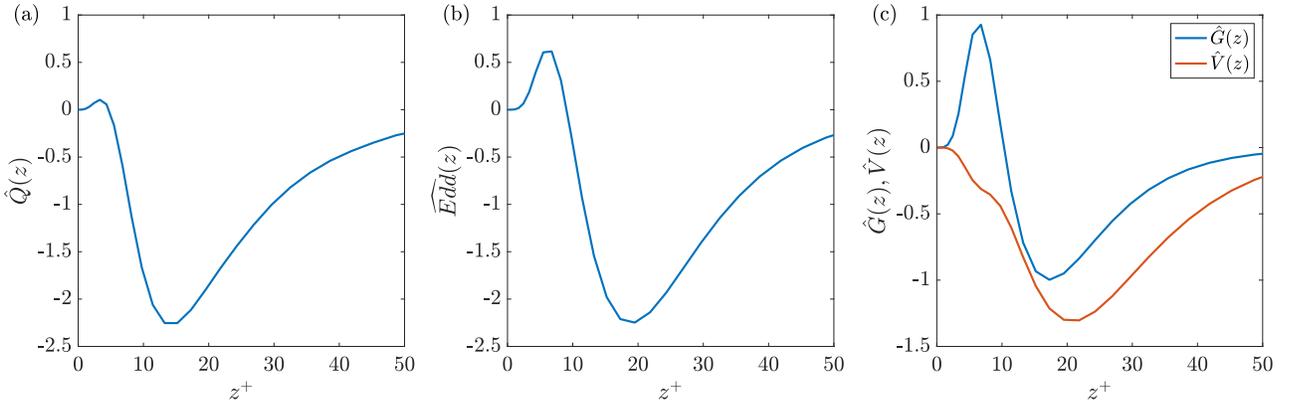}
\caption{(a) Time-averaged nonlinear transfer from DNS for $(k_x,k_y) = (4,30)$ compared to (b) eddy dissipation and (c) its components for $c^+ = 10$.} \label{fig:edd nwc}
\end{figure}

Figure~\ref{fig:edd nwc}(a) compares $\hat{Q}(z)$ for $(k_x,k_y) = (4,30)$ with $\widehat{Edd}(z)$ in Fig.~\ref{fig:edd nwc}(b). The eddy predictions are for the most energetic wave speed $c^+ = 10$. Good agreement can be observed between $\hat{Q}(z)$ and $\widehat{Edd}(z)$ although the eddy dissipation is most negative at $z^+ = 20$ instead of $z^+ = 15$. Similar to the previous scale, eddy dissipation overestimates the positive energy transfer in the near-wall region which, as seen in Fig.~\ref{fig:edd nwc}(c), is driven by $\hat{G}(z)$. The contribution from $\hat{V}(z)$, on the other hand, is negative at all wall-normal locations. It can be concluded that the eddy viscosity gradient plays an important role in modeling positive energy transfer processes. The predicted $\widehat{Edd}(z)$, nonetheless, exceeds the true nonlinear transfer in the near-wall region. The next section considers the effect of artificially weakening the eddy viscosity gradient to analyze its impact on the resulting mode shapes. 

\subsection{Artificially adjusting the eddy viscosity gradient} \label{sec:DnuT}

\begin{figure} 
\centering
\includegraphics[trim=3.9cm 0cm 0cm 0cm, clip=true, scale=0.4]{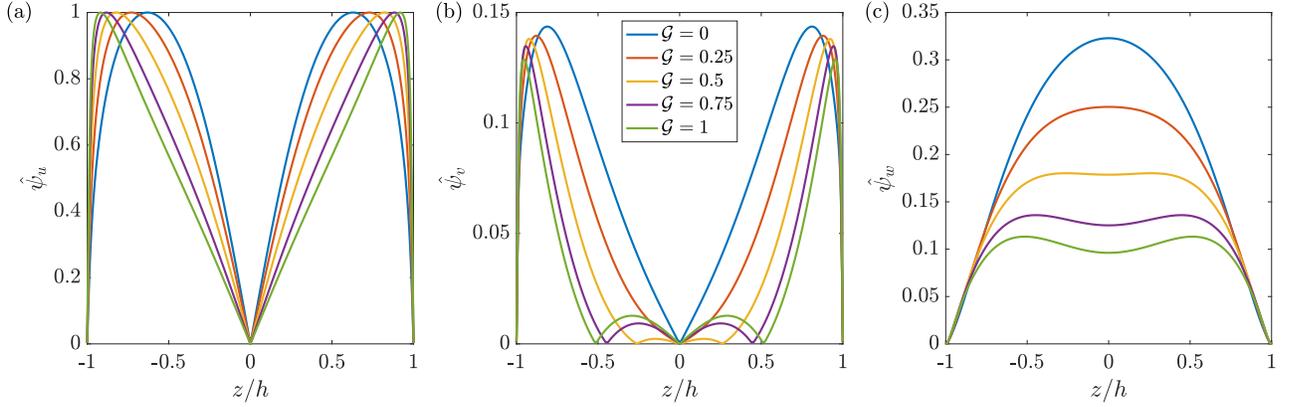}
\caption{Leading eddy modes for $\boldsymbol{k} = (0,4,0)$ for various strengths of the eddy viscosity gradient $\mathcal{G}$.} \label{fig:grad04}
\end{figure}

In this section, the strength of the eddy viscosity term is altered by introducing a scaling factor $\mathcal{G}$ such that Eq.~(\ref{eq:viscous index}) becomes
\begin{equation} \label{eq:Ggradient}
\boldsymbol{\nabla} \cdot \left[\nu_T(\boldsymbol{\nabla} \hat{\boldsymbol{u}} + \boldsymbol{\nabla} \hat{\boldsymbol{u}}^T) \right] = \nu_T \boldsymbol{\nabla}^2 \hat{\boldsymbol{u}} + \mathcal{G} (\boldsymbol{\nabla} \cdot \nu_T) (\boldsymbol{\nabla} \hat{\boldsymbol{u}} + \boldsymbol{\nabla}\hat{\boldsymbol{u}}^T),
\end{equation}
where $\mathcal{G} \geq 0$ controls the strength of the eddy viscosity gradient term. Figure~\ref{fig:grad04} illustrates the impact of artificially adjusting $\mathcal{G}$ for Mode 0. It can be seen that as $\mathcal{G}$ increases, the peak energy of the streamwise and spanwise velocity components shifts closer to the wall. The shape of the wall-normal velocity component is less affected but its magnitude decreases. It can be reasoned that increasing $\mathcal{G}$ results in greater positive energy transfer in the near-wall region. The energy of the modes, consequently, is redistributed towards the wall and to the wall-parallel velocity components. For some scales, $\mathcal{G}$ is too strong, resulting in too much energy in the streamwise and spanwise velocity components near the wall as seen for Mode 2 in Sec.~\ref{sec:mode structures}. These biased mode shapes explain why linear-based estimation techniques that use an eddy viscosity model, e.g. Refs.~\cite{Madhusudanan19,Gupta21}, overpredict the strength of fluctuations in the near-wall region when measurements are known in the logarithmic region. The biased mode shapes are also consistent with Ref.~\cite{Amaral21} who found that the eddy viscosity model underestimates fluctuations when using wall-based measurements. 

\begin{figure} 
\centering
\includegraphics[trim=4cm 0cm 0cm 0cm, clip=true, scale=0.4]{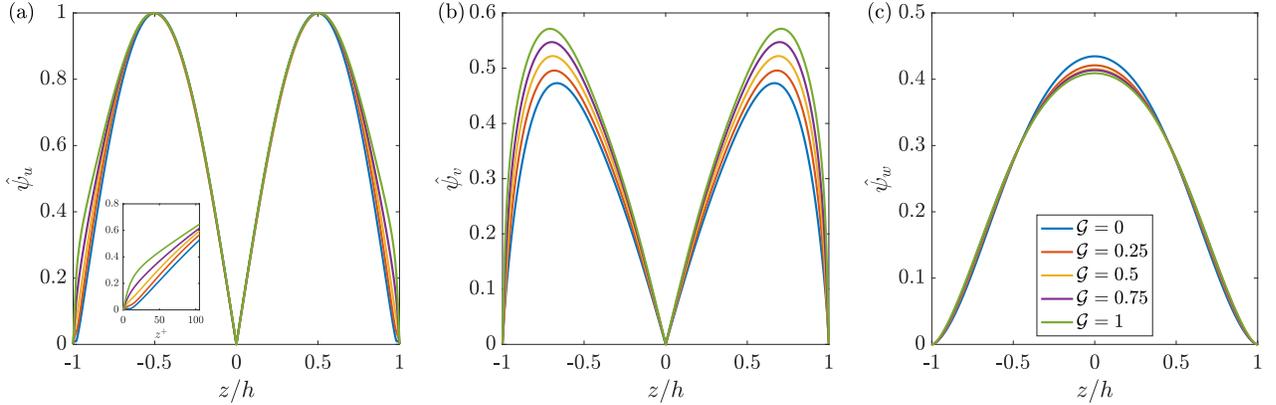}
\caption{Leading eddy modes for $\boldsymbol{k} = (1,2,18.5)$ for various strengths of the eddy viscosity gradient $\mathcal{G}$.} \label{fig:grad12}
\end{figure}

Figure~\ref{fig:grad12} presents the effect of $\mathcal{G}$ on the Mode 1 shapes. Unlike Mode 0, $\mathcal{G}$ has almost no impact on the structures although the spanwise component becomes slightly less energetic relative to the other velocity components as $\mathcal{G}$ increases. The main difference occurs in the streamwise velocity component very close to the wall. The inset of Fig.~\ref{fig:grad04} shows that the mode shape becomes flatter or more blunt as $\mathcal{G}$ increases. Similar to Mode 0, the flatter profiles can be attributed to $\hat{G}$ injecting energy in this region of the flow. This also explains why the constant eddy viscosity model, which has a wall-normal gradient of zero everywhere, was not able to reproduce the near-wall behavior in Fig.~\ref{fig:constantRe}(c) that appeared in Fig.~\ref{fig:constantRe}(b).

\subsection{Discussion} \label{sec:discussion}

The previous sections have analyzed the energy transfer processes that can be modeled by the Cess eddy viscosity profile. This section aims to contextualize these results with respect to recent low-order modeling efforts in the literature. One key challenge is to identify the smallest subset of nonlinear interactions that are needed to sustain a wall-bounded turbulent flow at high Reynolds numbers. The generalized quasi-linear (GQL) approximation \cite{Marston16}, in particular, is able to reproduce key statistical features of wall-bounded turbulence \cite{Hernandez21b}. The GQL approximation decomposes the flow into a low-wavenumber group and a high-wavenumber group. Nonlinear interactions involving the high-wavenumber group are removed. If the low-wavenumber group is restricted to the mean flow, the quasi-linear approximation (QLA) is recovered \cite{Farrell07,Marston08,Thomas14,Thomas15}. 

Reference~\cite{Hernandez21b} has shown that the GQL approach retains triadic interactions that are responsible for the scattering mechanism and inverse energy transfer in the near-wall region. The former can be attributed to low-high wavenumber interactions that feed into high wavenumbers while the latter arise from high-high wavenumber interactions that feed into low wavenumbers. It is posited that the negative energy transfer modeled by eddy viscosity through $\hat{V}$ can be interpreted as a scattering mechanism that removes energy from larger scales and redistributes it to smaller scales. The positive energy transfer modeled by $\hat{G}$, on the other hand, reproduces the inverse energy transfer in the near-wall region. The combined effect of these two transfers, therefore, encapsulates the effect of small scales on the large scales. 

\section{Conclusions} \label{sec:conclusions}

The predictions of resolvent analysis with and without eddy viscosity have been evaluated for a friction Reynolds number of $Re_\tau = 550$. The accuracy of the predictions were assessed using scalar measures including low-rank maps of the operator, projection coefficients, and the most amplified wave speed as well as direct comparison of the structures. The addition of eddy viscosity distorted the linear amplification mechanisms that are identified by resolvent analysis. As such, the low-rank maps of the eddy operator did not align well with the turbulent kinetic energy spectrum in comparison to the low-rank maps of the standard resolvent operator. For eddy analysis, low-rank behavior was identified for structures with spanwise wavelengths of $\lambda_y^+ \approx 80$ and $\lambda_y/h = 3.5$ independent of the wave speed under investigation. This is consistent with previous transient growth and optimal harmonic forcing studies \cite{delAlamo06,Hwang10}. 

Despite the distorted low-rank maps, the predictions from eddy analysis were generally in better agreement with DNS in comparison to resolvent analysis. The highest projection coefficients were obtained for eddy modes with wave speeds of $c^+ = 10$ and $c^+ = 18.9$. These correspond to structures associated with the near-wall cycle and structures that are most energetic at $z/h = \pm0.5$. The most amplified wave speed identified by eddy analysis was also found to match the most energetic wave speed in DNS for these two types of structures. For most wavenumber pairs, however, both resolvent and eddy analysis failed to predict the most energetic wave speed. Resolvent analysis overestimated the correct wave speed while eddy analysis underestimated the correct wave speed. 

The importance of wave speed on the projection coefficients motivated the selection of specific scales for comparing the SPOD modes to the resolvent and eddy modes. Consistent with previous studies, resolvent modes tended to be localized around the critical layer \cite{McKeon10} and the streamwise velocity component was too strong relative to the spanwise and wall-normal velocity components. All three velocity components of the eddy modes, meanwhile, matched their SPOD counterparts for the wave speeds of $c^+ = 10$ and $c^+ = 18.9$. For other wave speeds, eddy analysis mostly underestimated the most energetic wave speed and predicted structures that were energetic closer to the wall than the SPOD modes. It was concluded that for higher Reynolds numbers, SPOD and eddy modes will agree for $c^+ = 10$ and $c^+ = U_{CL}^+-2$ due to the invariance of the Cess profile in the near-wall region up to $z^+ = 15$, when it is premultiplied by $Re_\tau$, and in the outer region where it is maximum at $z/h = \pm 0.5$. Modes for $c^+ = U_{CL}^+-2$ could also be obtained with reasonable accuracy from eddy analysis by setting constant $\nu_T = Re_T = 12.5$.

Finally, there was an investigation into the energy processes that are modeled by the Cess profile. Two transfers were identified, one of which is an effective Reynolds number that varies spatially. Similar to dissipation, the effective Reynolds number term is guaranteed to be real and negative. The second term originates from the eddy viscosity gradient, which tends to be positive in the near-wall region. The combined effect, called eddy dissipation, models well the actual nonlinear transfer taking place in DNS although the positive energy transfer near the wall is overestimated. This explains the tendency for eddy modes to be too energetic near the wall as observed for higher Reynolds numbers in Ref.~\cite{Symon20} and in linear-based estimation, e.g. Ref.~\cite{Madhusudanan19}. 

In order to further improve predictions from resolvent analysis, the Cess profile could be replaced with a scale-dependent eddy viscosity. Reference~\cite{Pickering21}, for example, solved an inverse problem to find the eddy viscosity profile that maximized the projection of the leading resolvent mode onto the leading SPOD mode. The drawback of this approach is that it requires sufficient data in order to compute the leading SPOD mode. As such, other approaches have modified the Cess profile using scaling arguments \cite{Gupta21} or a stochastic approach to model background turbulence that can inject or dissipate energy of coherent waves \cite{Tissot21}. Regardless of the approach selected, the Cess profile provides a good initial condition for optimizing an eddy viscosity or modeling the effect of nonlinear terms as done in Ref.~\cite{Zare17}.    

\section*{Acknowledgments}

The authors would like to acknowledge the use of the IRIDIS High Performance Computing Facility, and associated support services at the University of Southampton. 

\appendix

\section{Linear operators} \label{sec:operators}

After elimination of the pressure, the linearized Navier-Stokes equations can be rewritten for the wall-normal velocity $\hat{w}$ and wall-normal vorticity $\hat{\eta} = ik_y\hat{u}-ik_x\hat{v}$. The matrices $\boldsymbol{A}$, $\boldsymbol{B}$, and $\boldsymbol{C}$ that appear in (\ref{eq:OSSQ}) are
\begin{subequations}
	\begin{equation}
	\boldsymbol{A} = \boldsymbol{M}\left[\begin{array}{cc} \mathcal{L}_{OS} & 0 \\ -ik_y U' & \mathcal{L}_{SQ} \end{array} \right],
	\end{equation}
	\begin{equation}
	\boldsymbol{B} = \boldsymbol{M} \left[\begin{array}{ccc} -i k_x \mathcal{D} & -i k_y \mathcal{D} & -k^2 \\ ik_y & -ik_x & 0 \end{array} \right],
	\end{equation}
	\begin{equation}
	\boldsymbol{C} = \frac{1}{k^2}\left[\begin{array}{cc} ik_x \mathcal{D} & -ik_y \\ ik_y \mathcal{D} & ik_x \\ k^2 & 0 \end{array} \right].
	\end{equation}
\end{subequations}
Both $\mathcal{D}$ and $'$ represent differentiation in the wall-normal direction and $k^2 = k_x^2+ k_y^2$. The mass matrix $\boldsymbol{M}$ is defined as
\begin{equation}
\boldsymbol{M}(k_x,k_y) = \left[\begin{array}{cc} \Delta^{-1} & 0 \\ 0 & \boldsymbol{I} \end{array} \right],
\end{equation}
where $\Delta = \mathcal{D}^2-k^2$ and $\boldsymbol{I}$ is the identity matrix. The Orr-Sommerfeld $\mathcal{L}_{OS}$ and Squire $\mathcal{L}_{SQ}$ operators are 
\begin{subequations}
	\begin{equation}
	\mathcal{L}_{OS} = -ik_xU\Delta + ik_xU'' + (1/Re_{\tau})\Delta^2,
	\end{equation}
	\begin{equation}
	\mathcal{L}_{SQ} = -ik_xU + (1/Re_{\tau}) \Delta.
	\end{equation}
\end{subequations}
With the addition of eddy viscosity, they become
\begin{subequations}
	\begin{equation}
	\mathcal{L}_{OS} = -ik_xU\Delta + ik_xU'' + \nu_T\Delta^2 + 2 \nu_T'\mathcal{D}\Delta + \nu_T''(\mathcal{D}^2 + k^2),
	\end{equation}
	\begin{equation}
	\mathcal{L}_{SQ} = -ik_xU + \nu_T \Delta + \nu_T'\mathcal{D}.
	\end{equation}
\end{subequations}

\bibliography{eddy}

\end{document}